\documentclass[11pt,a4paper]{article}
\usepackage[margin=1.25in]{geometry}                
\usepackage{amssymb,amsmath,xparse,amsfonts}
\usepackage{hyperref}
\usepackage{graphicx}

\usepackage{amsthm}
\usepackage{subcaption}
\usepackage{siunitx}
\usepackage{todonotes}

\usepackage{bm}

\newcommand{\tr}{\top}

\newcommand{\yp}[1]{{\color{black}{#1}}}
\newcommand{\ypb}[1]{{\color{black}{#1}}}

\newcommand{\ve}{\varepsilon}
\newcommand{\pa}{\partial}

\newcommand{\vc}{\text{vec}}

\newcommand{\dd}{\,\mathrm{d}}
\newcommand{\ies}{\int_0^\infty e^{\kappa_i s} }

\newcommand{\mi}{\mathrm{i}} 

{
	\theoremstyle{plain}
	\newtheorem{assumption}{Assumption}
}

\title{$N$-Body Oscillator Interactions of Higher-Order Coupling Functions}
\author{Youngmin Park and Dan Wilson}

\begin{document}
    \maketitle


\begin{abstract}
    We introduce a method to identify phase equations that include $N$-body interactions for general coupled oscillators valid far beyond the weak coupling approximation. This strategy is an extension of the theory from [Park and Wilson, SIADS 20.3 (2021)] and yields coupling functions for $N\geq2$ oscillators for arbitrary types of coupling (e.g., diffusive, gap-junction, chemical synaptic). These coupling functions enable the study of oscillator networks in terms of phase-locked states, whose stability can be determined using straightforward linear stability arguments. We demonstrate the utility of our approach with two examples. First, we use a diffusely coupled complex Ginzburg-Landau (CGL) model with $N=3$ and show that the loss of stability in its splay state occurs through a Hopf bifurcation {viewing the non-weak diffusive coupling as the bifurcation parameter. Our reduction also captures asymptotic limit-cycle dynamics in the phase differences}. Second, we use $N=3$ realistic conductance-based thalamic neuron models and show that our method correctly predicts a loss in stability of a splay state for non-weak synaptic coupling. In both examples, our theory accurately captures model behaviors that weak and recent non-weak coupling theories cannot.
\end{abstract}

\section{Introduction}

Oscillatory phenomena exist in many biological \cite{butera1999models,winfree2001geometry,strogatz2005crowd,izhikevich2007,Zang_Marder_2023}, chemical \cite{kuramoto84,epstein1998introduction}, and physical systems \cite{ott2008low}. Numerical models that capture the important behaviors of these systems often involve complex interactions of large numbers of variables, reducing the visibility of important mechanisms.  As such, phase reduction is often used for understanding the aggregate behavior of interacting oscillators in a reduced order setting  \cite{kuramoto84,izhikevich2007,ermentrout2010,park2017utility,ermentrout2019recent}.

The many techniques developed for phase reduction often include specific assumptions that improve tractability at the cost of biological relevance. The Kuramoto model is an exceptionally well-studied model, owing to its elegant simplicity, and has proven invaluable towards understanding higher-order interactions and stable synchronous network states \cite{leon2019phase}. However, the Kuramoto model was originally derived in the case of infinite, homogeneous, globally coupled oscillators \cite{da2018kuramoto} near the Hopf bifurcation \cite{leon2022enlarged}, limiting its use for finite populations away from the Hopf. Moreover, its analysis is often limited to studying questions around synchrony as opposed to other phase-locking phenomena, due to the often-taken limit of infinite oscillators.

When a finite number of oscillators is considered, other features may be exploited, each with their own limitations. When the network exhibits symmetries, it is possible to enumerate all phase-locked states with weak or strong coupling \cite{golubitsky2016symmetry}, but this method is not suited to work in the case of asymmetries \cite{hunter2022pattern}. In networks of neurons, the pulse-like shape of action potentials allows for the use of pulse coupling \cite{cui2009functional, canavier2009phase,canavier2010pulse,peskin1975mathematical,mirollo1990synchronization}. This approach yields analytically tractable results for weak or strong and possibly asymmetric coupling, but the number of oscillators is often limited to pairs. The study of network behavior can be made tractable by using piecewise smooth models, but coupling functions require particular assumptions such as linear coupling \cite{coombes2012nonsmooth,coombes2018networks}, weak coupling \cite{coombes2008neuronal,park2018infinitesimal}, and Laplacian coupling \cite{nicks2018clusters}. In addition, the analysis of phase-locked states is often restricted to understanding the stability of a {synchronous network state \cite{coombes2018networks,coraggio2021convergence} (although some do consider the stability of splay states \cite{coombes2008neuronal}).}

The most relevant reduction for the present study is the theory of weakly coupled oscillators, which allows for a general form of the vector field and coupling function so long as the coupling strength is weak \cite{ermentrout1984frequency,kopell1986symmetry,park2016weakly,park2017utility,park2018infinitesimal,park2018multiple}. \yp{To be more precise, by \textit{weak} coupling, we mean phase reductions that only consider expansions up to first order in coupling strength (often represented by $\ve$), and are thus generally only guaranteed to be valid for arbitrarily small $\ve$. } The weak assumption is a severe limitation because it cannot necessarily be used to accurately capture the dynamical behavior of coupled oscillator networks in many biological networks, e.g., cortical networks \cite{pfeffer2013inhibition,chaudhuri2015large}, subcortical networks \cite{thibeault2013using}, and pacemaker networks \cite{butera1999models,golowasch1992contribution}. Indeed, recent studies have pushed beyond the weak coupling regime by deriving correction terms in higher orders of the coupling strength \yp{(i.e., non-weak coupling)}, but these too have their limitations. Higher order phase correction terms considered in \cite{rosenblum2019numerical,genge2020high,von2023parametrisation} require the underlying limit cycle to be \textit{strongly attracting}, limiting their applicability when Floquet multipliers are close to unity \cite{wilson2018greater}. Recently developed isostable coordinates have proven invaluable towards developing more robust phase reductions, e.g., \cite{wilson2019phase,park2021high,nicks2024insights}. However, these methods have only been applied to pairs of oscillators without heterogeneity \ypb{(except in \cite{nicks2024insights}, where the authors consider the complex Ginzburg-Landau model for $N\geq2$ and the Morris-Lecar model for $N=200$)}, and \yp{a recently-published article by \cite{mau2023high} closely mirrors our assumptions, but is only valid for planar systems}. \ypb{The recent work by Nicks et al \cite{nicks2024insights} is of significant relevance to this paper, and we briefly contrast our results in the Discussion (Section \ref{sec:discussion}).} 

In networks consisting of more than 2 oscillators, $N$-body interactions on simplicial complexes become relevant. Much recent work has been done to develop phase reductions in this direction, but the study of $N$-body interactions have been limited to \yp{tractable} models such as the Kuramoto model \cite{skardal2011cluster,tanaka2011multistable,skardal2020higher,bick2022multi} or the Ginzbug-Landau equation \cite{leon2019phase}. Finally, note that these studies begin with higher-order interactions as an assumption, in contrast to \cite{wilson2019phase}, where it is shown that higher-order interactions emerge as a function of higher-order corrections to weak coupling theory. Thus, there exists no phase-isostable reduction method for general $n$-dimensional oscillators that accounts for heterogeneity, strong coupling, and resulting $N$-body interactions.

In the present study, we address this gap in the literature by deriving a phase reduction method applicable to networks of (weakly or strongly attracting) coupled oscillators with arbitrary network topology beyond weak coupling, \yp{i.e., we calculate higher-order corrections to the first-order reduction methods while incorporating isostable coordinate(s)}. The formulation includes $N$-body interactions on simplicial complexes and enables us to study the existence and stability of phase-locked states in a manner not possible using the original models.

The paper is organized as follows. We derive the proposed phase-isostable reduction in Section \ref{sec:nonweak} and check the performance of our method using the complex Ginzburg-Landau (CGL) model (with some analytical calculations) in Section \ref{sec:cgl}. We also confirm that our method works using the neurobiologically realistic thalamic model in Section \ref{sec:thal}. Stability of splay states as a function of coupling strength are discussed as part of these results. We conclude the paper with a discussion in Section \ref{sec:discussion}.

All code used to generate figures are available for public use at \url{https://github.com/youngmp/nbody} \cite{parkgithub}.

\section{Background}

\subsection{Phase and Phase Reduction}
Consider a general dynamical system
\begin{equation} \label{maineq}
    \dot{X} = F(X) + U(X,t),
\end{equation}
where $X \in \mathbb{R}^n$ is the state, $F:\mathbb{R}^n \rightarrow \mathbb{R}^n$ is a smooth vector field, and $U(X,t) \in \mathbb{R}^n$ is some additive input.  Let $Y$ be a stable $T$-periodic orbit that emerges when taking $U(X,t) = 0$. In situations where the timing of oscillations is of interest, it can be useful to consider the dynamics of Equation \eqref{maineq} in terms of a scalar phase $\theta(X) \in \mathbb{S}^1$ rather than in terms of the state.  When $U(X,t) = 0$, the notion of isochrons \cite{guck75} can be used to define phase in the basin of attraction of the limit cycle.  Isochrons can be defined as follows:~letting $\theta_1 \in [0,1)$ be the phase associated with an initial condition $a(0) \in Y$, the $\theta_1$ isochron is comprised of the set of all $b(0)$ for which
\begin{equation} \label{isodef}
    \lim_{t \rightarrow \infty} || a(t)-b(t)||=0,
\end{equation}
where $||\cdot||$ can be any vector norm.  Isochrons are typically scaled so that $\frac{d \theta}{dt}$ is a constant for trajectories that evolve under the unperturbed flow of the vector field; in this work, we choose $\frac{d \theta}{dt} = 1$ for convenience.

Working in phase coordinates, by restricting ones attention to a close neighborhood of the periodic orbit and allowing $U(X,t)\neq0$, through a change of coordinates one arrives at the standard phase reduction \cite{ermentrout2010,izhikevich2007,kuramoto84}
\begin{align} \label{predeq}
    \frac{d \theta}{dt} &= \frac{\partial \theta}{\partial X} \cdot \frac{dX}{dt} \nonumber \\
    &= \frac{\partial \theta}{\partial X} \cdot \big( F(X) + U(X,t) \big) \nonumber \\
    &= 1 + \mathcal{Z}(\theta) \cdot U(X,t),
\end{align}
where $\mathcal{Z}(\theta) = \frac{\partial \theta}{\partial X}$ evaluated on the periodic orbit at phase $\theta$, and the `dot' denotes the dot product.  In the third line above, we use the fact that $\frac{\partial \theta}{\partial X} \cdot F(X)$ was scaled to equal 1.  Reductions of the form \eqref{predeq} have been used widely a starting point for the analysis of weakly perturbed and weakly coupled oscillatory systems \cite{ermentrout2010,piet19,mong19,wils22review}.

\subsection{Isostable Coordinates}
In Equation \eqref{predeq}, the gradient of the phase is evaluated on the periodic orbit.  As such, it requires the state $X$ to remain close to $Y$ for the reduction to remain valid.  This is only guaranteed in the limit of weak forcing; in many practical applications, alternative techniques that can accommodate stronger forcing must be used.  One common strategy is to augment the phase coordinates with amplitude coordinates.  A variety of techniques for considering both phase and amplitude coordinates have been proposed \cite{wils16isos,rosenblum2019numerical,wedg13,wils18operat,shir17}.

In this work, we  use the phase-isostable coordinate system, which augments the phase dynamics with additional information about level sets of the slowest decaying modes of the Koopman operator \cite{maur13,mezi20}.  To illustrate this coordinate system, let $U(X,t) = 0$ and define $\Delta X = X - Y(\theta)$.  To a linear approximation the dynamics of Equation \eqref{maineq} follow
\begin{equation} \label{deltaxeq}
    \Delta \dot{X} = J \Delta X,
\end{equation}
where $J$ is the Jacobian evaluated at $Y(\theta(t))$.  Notice that \eqref{deltaxeq} is linear and time-varying with the Jacobian being $T$-periodic.  Let $\Phi$ be the fundamental matrix, i.e.,~with  $\Delta X(T) = \Phi \Delta X(0)$ for initial conditions $\theta(X(0)) \approx 0$.  Further, let $w_j, v_j$, and $\lambda_j$ be left eigenvectors, right eigenvectors, and associated eigenvalues, respectively, of $\Phi$.  Floquet exponents can be computed according to $\kappa_j = \log(\lambda_j)/T$.  Let $\kappa_1$ be the slowest decaying nonzero Floquet exponent. If $\kappa_1$ is unique, an associated isostable coordinate can be defined in the basin of attraction of the limit cycle according to \cite{wilson2018greater}
\begin{equation} \label{isostabledef}
    \psi_1(X) = \lim_{k \rightarrow \infty}(w_1^\tr(\ypb{\eta}(t^k_\Gamma,X) - Y_0   )   \exp(-\kappa_1 t_\Gamma^k)),
\end{equation}
where $t_\Gamma^k$ denotes the time of the $k$th transversal of the $\theta = 0$ isochron, $\eta(t,X)$ is the unperturbed flow of the vector field that takes $X(0)$ to $X(t)$,  $Y_0$ is the intersection of the periodic orbit and the $\theta = 0$ isochron, and $^\tr$ denotes the transpose.  In contrast to isochrons defined in \eqref{isodef}, which characterize the infinite time convergence to the periodic orbit, the isostable coordinates defined in \eqref{isostabledef} give a sense of the distance from the periodic orbit, with larger $|\psi_1(X)|$ values corresponding to states that will take longer to approach the periodic orbit.  Isostable coordinates can also be used to characterize faster decaying components of the solution, but an explicit definition akin to \eqref{isostabledef} is not always possible \cite{kval21}. Instead, faster decaying isostable coordinates can be defined as level sets of appropriately chosen Koopman eigenfunctions. In this work, we will assume that the faster decaying isostable coordinates decay rapidly and are well-approximated by zero.  Using Equation \eqref{isostabledef}, it is possible to show directly that when $U(X,t) = 0$, $\frac{d \psi_1}{dt} = \kappa_1 \psi_1$ in the basin of attraction of the limit cycle \cite{wilson2018greater}.

\subsection{Phase-Isostable Reduction}
Information about the slowest decaying isostable coordinate can be used to augment standard phase models of the form \eqref{predeq} to increase the accuracy of the reduction in response to larger magnitude inputs.  In the analysis in the following sections, we assume that all non-zero Floquet exponents except $\kappa_1$ have a large real component so that the associated isostable coordinates decay rapidly and are well-approximated by zero.  Moving forward, for notational convenience, we will simply use $\psi$ and $\kappa$ to denote the only non-truncated isostable coordinate and its Floquet exponent -- \yp{\textit{from this point forward, subscripts of any variable including $\psi$ and $\kappa$ will only denote the oscillator index}}.  Taking this isostable coordinate into account, one can consider a modified version of \eqref{predeq}
\begin{equation} \label{phaseisoeq}
    \dot{\theta} = 1 + \mathcal{Z}(\theta,\psi) \cdot U(X,t),
\end{equation}
where the gradient of the phase is not necessarily evaluated on the periodic orbit, but rather, at a state corresponding to $X(\theta,\psi)$; note that $X(\theta,0) = Y(\theta)$.  In order to use \eqref{phaseisoeq}, it is necessary to consider the isostable coordinate dynamics as well as the phase dynamics.  Considering the dynamics given by Equation \eqref{maineq}, by the chain rule,
\begin{align} \label{isoeq}
    \frac{d \psi}{dt} &= \frac{\partial \psi}{\partial X} \cdot \frac{d\ypb{X}}{dt} \nonumber \\
    &= \frac{\partial \psi}{\partial X} \cdot (F(\ypb{X}) + U(X,t)) \nonumber \\
    &= \kappa \psi + \mathcal{I}(\theta,\psi) \cdot U(X,t),
\end{align}
where $\mathcal{I}(\theta,\psi) = \frac{\partial \psi}{\partial X}$ evaluated at $X(\theta,\psi)$.  In the third line above, the relationship $\frac{\partial \psi}{\partial X} \cdot F(x) = \kappa \psi$ since $\frac{d \psi}{dt} = \kappa \psi$ when $U(X,t)=0$.  Taken together, Equations \eqref{phaseisoeq} and \eqref{isoeq} comprise the phase-isostable reduction.  For computation and analysis purposes, the gradient of the phase and isostable coordinate is often represented according to a Taylor expansion in the isostable coordinate centered at $\psi = 0$
\begin{align} \label{phaseampred}
    \dot{\theta} &= 1 + (Z^{(0)}(\theta) + \psi Z^{(1)}(\theta) + \psi^2 Z^{(2)}(\theta) + \dots) \cdot U(X,t), \nonumber \\
    \dot{\psi} &= \kappa \psi + (I^{(0)}(\theta) + \psi I^{(1)}(\theta) + \psi^2 I^{(2)}(\theta) + \dots) \cdot U(X,t),
\end{align}

\section{Higher Order Coupling with N-Body Interactions}\label{sec:nonweak}

We now derive a reduced system of phase equations that captures  higher-order interactions between coupled oscillators, starting with the ordinary differential equation (ODE)
\begin{equation}\label{eq:odes}
    \dot X_i = F_i(X_i) + \delta_i \yp{Q}_i(X_i) + \ve \left[ \sum_{j=1}^N a_{ij} G_{ij}(X_i,X_j) \right], \quad i=1,2,\ldots,N,\\
\end{equation}
where each system admits a $T$-periodic limit cycle $Y_i(t)$ when $\delta_i=\ve=0$.  \yp{We assume, when $\delta_i,\ve\neq0$, that $\delta_i$ and $\ve$ are not necessarily small, and that $\delta_i$ is order $O(\ve)$. Thus, while the derivation to follow is constrained by $\ve$, heterogeneity and coupling need not be weak in the sense that we seek to capture higher-order corrections to first-order phase dynamics.}

We assume general smooth vector fields $F_i:\mathbb{R}^{n_i} \rightarrow \mathbb{R}^{n_i}$, smooth coupling functions $G_{ij}:\mathbb{R}^{n_i}\times\mathbb{R}^{n_j}\rightarrow \mathbb{R}^{n_i}$, and smooth additive heterogeneity $Q_i:\mathbb{R}^{n_i} \rightarrow \mathbb{R}^{n_i}$, where $n_i \in \mathbb{N}$ for each oscillator $i$. The scalars $a_{ij}$ modulate coupling strength between pairs of oscillators \yp{and determine network topology}, whereas $\ve$ modulates the overall coupling strength of the network.

 \yp{\paragraph{Remarks}
    \begin{itemize}
        \item \ypb{Oscillators need not necessarily share the same period to apply the proposed method. Requiring all oscillators to have the same period when $\delta_i=\ve=0$ is a matter of convenience, but equivalent to heterogeneity in the following sense. The dynamics of node $i$ without coupling ($\ve=0$) are given by $\dot X_i = F_i(X_i,c_i^*) + \delta_i Q_i(X_i)$, where $c_i^*$ is a particular choice of a parameter in $F_i$ such that when $\delta_i=0$, the stable limit cycle of the node dynamics $Y_i(t,c_i^*)$ has period $T$ for each $i$. While we require that $|c_i-c_i^*| = O(\ve)$ for all $i$, we remind the reader that $\ve$ is not necessarily small -- and our method is designed to capture potential higher-order effects. This property is the same as that considered in \cite{mau2023high} -- we save a more detailed comparison for the Discussion (Section \ref{sec:discussion}). As a natural example, it is reasonable to expect such $c_i^*$ to exist in multiple heterogeneous neural oscillators by adjusting their individual input currents, with the caveat that the heterogeneity is not too large}. 
        \item The choice of heterogeneity in \eqref{eq:odes} implicitly restricts our attention to cases where the difference $c_i - c_i^*=\delta_i$ yields a linear change in each vector field. However, nonlinear changes may be incorporated by including additional higher-order terms in $\delta_i$ (we elaborate more on this feature at the end of the phase difference derivation in Section \ref{sec:phase_diff}).
        \item \ypb{In principle, there may be any number of heterogeneous parameters per oscillator (as opposed to one $c_i^*$ per oscillator $i$)}, but we restrict our attention to the simple case of one heterogeneous parameter across all oscillators ($\delta_i \equiv \delta$) because the primary goal of the present study is to verify that our proposed phase reduction accurately captures higher-order interactions due to \textit{coupling}.
\end{itemize} }

We assume that there is only one nontrivial isostable coordinate similar to prior studies \cite{wilson2018greater,wilson2019phase,park2021high} and let $\yp{\kappa_i}<0$ be the corresponding Floquet exponent, \yp{where we place no restrictions on the spread of $\kappa_i$ between oscillators (we reiterate that the subscript $i$ denotes the oscillator index, so $\kappa_i$ is not the $i$th Floquet exponent of a given oscillator, but rather the Floquet exponent of the nontrivial isostable coordinate of oscillator $i$).}

We reduce \eqref{eq:odes} to phase-amplitude coordinates using phase-isostable reduction of the form \eqref{phaseisoeq} and \eqref{isoeq}:
\begin{equation}\label{eq:reduced0}
    \begin{split}
        \dot \theta_i &= 1+\ypb{\delta} \mathcal{Z}_i(\theta_i,\psi_i) \cdot Q_i(\theta_i,\psi_i)+ \ve\mathcal{Z}_i(\theta_i,\psi_i) \cdot \sum_{j=1}^N a_{ij} G_{ij}(\theta_i,\psi_i,\theta_j,\psi_j),\\
        \dot \psi_i &= \yp{\kappa_i} \psi_i + \ypb{\delta}\mathcal{I}_i(\theta_i,\psi_i)\cdot Q_i(\theta_i,\psi_i)+ \ve\mathcal{I}_i(\theta_i,\psi_i)\cdot \sum_{j=1}^N a_{ij} G_{ij}(\theta_i,\psi_i,\theta_j,\psi_j),
    \end{split}
\end{equation}
where $\theta_i$ represents the oscillator phase, $\psi_i$ represents the amplitude of a trajectory perturbed away from the underlying limit cycle, $\mathcal{Z}_i$ is the gradient of the phase often referred to as the phase response curve (PRC), and $\mathcal{I}_i$ is the gradient of the isostable coordinate often referred to as the isostable response curve (IRC). We suppress the time dependence of $\theta_i$ \yp{and $\psi_i$} to reduce notational clutter.

\subsection{Expansions}

\yp{An important step towards reducing \eqref{eq:odes} involves taking the Taylor expansion of all terms with respect to $\psi_i$ and $\ve$. Assuming that the functions $\mathcal{Z}_i$ and $\mathcal{I}_i$ are \yp{sufficiently} smooth within their respective basins of attraction}, their approximations may be obtained to \ypb{high degrees of accuracy} by computing the coefficients of $\psi_i$ as in Equation \eqref{isoeq} (see \cite{wilson2020phase,perez2020global}):
\begin{align}
    \mathcal{Z}_i(\theta,\psi) &\approx Z_i^{(0)}(\theta) + \psi Z_i^{(1)}(\theta) + \psi^2 Z_i^{(2)}(\theta) +\cdots,\label{eq:z_exp}\\
    \mathcal{I}_i(\theta,\psi) &\approx I_i^{(0)}(\theta) + \psi I_i^{(1)}(\theta) + \psi^2 I_i^{(2)}(\theta) +\cdots,\label{eq:i_exp}\\
    X_i(t) &\approx Y_i(\theta_i) + \psi_i g_i^{(1)}(\theta_i)+ \psi_i^2g_i^{(2)}(\theta_i)+\cdots,\label{eq:x_exp}
\end{align}
where $Z_i^{(k)}$, $I_i^{(k)}$, and $g_i^{(k)}$ are the higher-order correction terms to the infinitesimal (linear) PRC, infinitesimal (linear) IRC, and Floquet eigenfunction, respectively. \yp{We make an additional ansatz that each isostable coordinate $\psi_i$ can be expressed as order $O(\ve)$ deviations from $\psi_i=0$ due to coupling and heterogeneity, and that corrections to the order $O(\ve)$ term can be computed directly from higher-order Taylor expansions. Thus,
\begin{equation}\label{eq:psi_exp}
    \psi_i(t) \approx \ve p_i^{(1)}(t) + \ve^2 p_i^{(2)}(t) + \ve^3 p_i^{(3)}(t) + \cdots.
\end{equation}}
We assume that we have calculated solutions $Z_i^{(k)}$, $I_i^{(k)}$, and $g_i^{(k)}$ for each $i=1,\ldots,N$ and $k\in\mathbb{N}$, for instance, using methods described in \cite{wilson2020phase}.

\yp{Some care must be taken when obtaining $\ve$-expansions for the coupling functions $G_{ij}$} (in contrast to the $\ve$-expansion of $\mathcal{Z}_i$, $\mathcal{I}_i$, and $X_i$, which is a straightforward matter of substituting \eqref{eq:psi_exp} into \eqref{eq:z_exp}, \eqref{eq:i_exp}, and \eqref{eq:x_exp}, respectively). Let us fix a particular pair of oscillators $i$ and $j$. We use the Floquet eigenfunction approximation for each oscillator,
\begin{equation}\label{eq:efuns}
    \Delta x_i \approx \psi_i g_i^{(1)}(\theta_i) + \psi_i^2 g_i^{(2)}(\theta_i) + \cdots,
\end{equation}
where $\Delta x_i \equiv X_i(t)-Y_i(\theta_i(t))$ is the difference between the limit cycle $Y_i$ and trajectory $X_i$. $\Delta x_j$ has an identical expression in terms of $j$ instead of $i$. We view the coupling function as the map $G_{ij}:\mathbb{R}^{n_i + n_j}\rightarrow \mathbb{R}^{n_i}$, where $G_{ij}(\Xi_{ij})=\left[G_{ij,1}(\Xi_{ij}),G_{ij,2}(\Xi_{ij}),\ldots G_{ij,n}(\Xi_{ij})\right]^\tr \in\mathbb{R}^{n_i}$, $G_{ij,m}:\mathbb{R}^{n_i+n_j}\rightarrow \mathbb{R}$, and $\Xi_{ij} = [X_i^\tr,X_j^\tr]^\tr \in \mathbb{R}^{n_i+n_j}$, an $(n_i+n_j)\times 1 $ column vector. Define $\Lambda_{ij} = [Y_i(\theta_i)^\tr,Y_j(\theta_j)^\tr]^\tr\in \mathbb{R}^{n_i+n_j}$ and $\Delta \Xi_{ij} = [\Delta x_i^\tr,\Delta x_j^\tr]^\tr\in \mathbb{R}^{n_i+n_j}$. Both are $(n_i + n_j) \times 1$ column vectors, so that the relation $\Xi_{ij} = \Lambda_{ij} + \Delta \Xi_{ij}$ is well-defined.

We apply the standard definition of higher-order derivatives from  \cite{magnus2019matrix,wilson2020phase} to obtain the multivariate Taylor expansion of $G_{ij,m}$ for each $m=1,2,\ldots,n$:
\begin{align}\label{eq:g_raw}
    G_{ij,m}(\Lambda_{ij} + \Delta \Xi_{ij})&= G_{ij,m}(\Lambda_{ij})+ G_{ij,m}^{(1)}(\Lambda_{ij})\Delta \Xi_{ij}+ \sum_{k=2}^\infty \frac{1}{k!}\left[ \stackrel{k}{\otimes} \Delta \Xi_{ij}^\tr\right] \vc\left(G_{ij,m}^{(k)}(\Lambda_{ij})\right),
\end{align}
    \yp{where the ``$\vc$'' operator simply reshapes a matrix by stacking its columns, allowing us to avoid calculating high-dimensional tensors. For example, if an $\ypb{m\times n}$ matrix $A$ has columns $a_i$ for $i=1,\ldots,n$ for $a_i \in\mathbb{R}^m$, then $\vc(A)$ is the $mn\times 1$ column vector $(a_1^\tr,a_2^\tr,\ldots,a_n^\tr)^\tr$. If $A$ is a Jacobian matrix, taking partial derivatives of $\vc(A)$ yields a matrix, whereas taking partial derivatives of $A$ yields a tensor. Both contain the same information, but the partials of $\vc({A})$ are much easier to implement in a computer.} Temporarily treating $\Lambda_{ij}$ as a vector of dummy variables, higher-order derivatives of $G_{ij,m}$ are given by
\begin{equation}\label{eq:g_deriv}
    G_{ij,m}^{(k)}(\Lambda_{ij}) = \frac{\pa \vc\left(G_{ij,m}^{(k-1)}(\Lambda_{ij})\right)}{\pa \Lambda_{ij}^\tr} \in \mathbb{R}^{(n_i+n_j)^{(k-1)}\times (n_i+n_j)},\quad k\geq 1.
\end{equation}

We replace $\Delta \Xi_{ij}$ in  \eqref{eq:g_raw} with the Floquet eigenfunction expansions \eqref{eq:efuns} and replace each $\psi_i$ with its expansion \eqref{eq:psi_exp}. With these substitutions in place, we collect the expansion of $G_{ij}$ in powers of $\ve$. The notation becomes cumbersome, so we summarize this step by writing
\begin{equation}\label{eq:G_eps}
    \begin{split}
        G_{ij}(\theta_i,\psi_i,\theta_j,\psi_j)= &\, K_{ij}^{(0)}(\theta_i,\theta_j)\\
        &+ \ve K_{ij}^{(1)}\left(\theta_i,\theta_j,p_i^{(1)},p_j^{(1)}\right)\\
        &+ \ve^2 K_{ij}^{(2)}\left(\theta_i,\theta_j,p_i^{(1)},p_i^{(2)},p_j^{(1)},p_j^{(2)}\right)\\
        &+ \ve^3 K_{ij}^{(3)}\left(\theta_i,\theta_j,p_i^{(1)},p_i^{(2)},p_i^{(3)},p_j^{(1)},p_j^{(2)},p_j^{(3)}\right)\\
        &+ \cdots.
    \end{split}
\end{equation}
The $K^{(\ell)}$ functions contain the collected Floquet eigenfunctions and the \yp{partial derivatives} of $G_{ij}$ for each order $O(\ve^\ell)$. \yp{We compute more explicit forms of $K_{ij}^{(\ell)}$ in Appendix \ref{a:expansion}} for use in some explicit calculations in Section \ref{sec:cgl_explicit}. It is straightforward to verify using a symbolic package that the function $K^{(k)}$ only depends on terms $p_i^{(\ell)}$, $p_j^{(\ell)}$ for $\ell \leq k$. For additional details, we refer the reader to our code repository \cite{parkgithub}, where we implement Sympy to automate the collection of symbolic terms \cite{sympy}.

\yp{Note that the $\ve$-expansion of $Q_i$ is just as straightforward to calculate using the above method. We summarize the expansion of $Q_i$ by writing
    \begin{align*}
        Q_i(\theta_i,\psi_i) =& Q_i^{(0)}(\theta_i)+\ve Q_i^{(1)}\left(\theta_i,p_i^{(1)}\right)+\ve^2 Q_i^{(2)}\left(\theta_i,p_i^{(1)},p_i^{(2)}\right)\\
        &+\ve^3 Q_i^{(3)}\left(\theta_i,p_i^{(1)},p_i^{(2)},p_i^{(3)}\right)+\cdots.
\end{align*}
More explicit forms for $Q_{i}^{(\ell)}$ are also in Appendix \ref{a:expansion}, and have almost the same form as the expression for $K_{ij}^{(\ell_1,\ell_2)}$. These more explicit terms will likewise be used in the Section \ref{sec:cgl_explicit} calculations.}

\subsection{Elimination of Isostable Coordinates}

While we now have all the necessary expansions in $\ve$ to rewrite the phase-amplitude equations \eqref{eq:reduced0} in powers of $\ve$, there remain two variables for each oscillator: $\theta_i$ and $\psi_i$. Thus, some work remains to reduce the system to a single phase variable. We proceed with the method suggested by \cite{wilson2019phase,park2021high}, by deriving and solving linear equations for each term $p_i^{(k)}$ in the expansion of $\psi_i$ \eqref{eq:psi_exp}. \yp{We state two key assumptions before proceeding}:

\yp{\begin{assumption}
        There exists a sufficient separation of timescales between $\hat\theta_i$ and $\theta_i$ (where $\hat\theta_i = \theta_i - t$), such that $\hat\theta_i$ can be treated as constant in time integrals. To be more precise, we assume that the \textit{averaged} dynamics of $\hat\theta$ is relatively slow compared to $t$.
\end{assumption}

\begin{assumption}
     First-order averaging is sufficient to capture nonlinear effects well beyond first order in $\ve$. This assumption implies some upper bound on $\ve$, but we will utilize higher-order averaging from, e.g., \cite{llibre2014higher,maggia2020higher} in future work, if needed.
\end{assumption}}

We begin by subtracting the moving frame and letting $\hat\theta_i = \theta_i-t$. Then the phase-amplitude equations \eqref{eq:reduced0} become
\begin{align}
    \dot{\hat\theta}_i &= \yp{\ypb{\delta} \mathcal{Z}_i(\hat\theta_i+t,\psi_i) \cdot Q_i(\hat\theta_i+t,\psi_i)} +\ve \sum_{j=1}^N a_{ij} \mathcal{Z}_i(\hat\theta_i+t,\psi_i) \cdot G_{ij}(\hat\theta_i+t,\psi_i,\hat\theta_j+t,\psi_j),\label{eq:th_moving}\\
    \dot \psi_i &=\yp{\kappa_i} \psi_i + \yp{\ypb{\delta} \mathcal{I}_i(\hat\theta_i+t,\psi_i) \cdot Q_i(\hat\theta_i+t,\psi_i)} + \ve \sum_{j=1}^N a_{ij} \mathcal{I}_i(\hat\theta_i+t,\psi_i)\cdot G_{ij}(\hat\theta_i+t,\psi_i,\hat\theta_j+t,\psi_j).\label{eq:psi_moving}
\end{align}
Substituting the expansion $\psi_i(t) = \ve p_i^{(1)}(t) + \ve^2 p_i^{(2)}(t) + \ve^3 p_i^{(3)}(t) + \cdots,$ into \eqref{eq:psi_moving} yields a hierarchy of ODEs in powers of $\ve$. The left-hand side consists of straightforward time derivatives:
\begin{equation*}
    \dot\psi_i = \ve \frac{d}{dt}p_i^{(1)} + \ve^2 \frac{d}{dt}p_i^{(2)}+ \ve^3 \frac{d}{dt}p_i^{(3)} +\cdots,
\end{equation*}
and the right-hand side includes many cross-multiplication terms \yp{(not all $\psi_i$ are shown expanded in $\ve$ for brevity)}:
\begin{align*}
    \kappa_i \psi_i + &\yp{\ypb{\delta} \mathcal{I}_i(\hat\theta_i+t,\psi_i) \cdot Q_i(\hat\theta_i+t,\psi_i)}+ \ve\sum_{j=1}^N a_{ij}\mathcal{I}_i(\hat \theta_i+t,\psi_i)\cdot G_{ij}(\hat\theta_i+t,\psi_i,\hat\theta_j+t,\psi_j)\\
    &= \kappa_i\left[\ve p_i^{(1)}(t) + \ve^2 p_i^{(2)}(t)+\cdots\right]\\
    &\quad +\yp{\ypb{\delta} \left\{\left[ I_i^{(0)}(\hat\theta_i+t) + \psi_i I_i^{(1)}(\hat\theta_i+t) + \psi_i^2 I_i^{(2)}(\hat\theta_i+t) +\cdots\right]\right.}\\
    &\quad\quad\quad\quad\yp{\cdot\left[ Q_i^{(0)}(\theta_i)+\ve Q_i^{(1)}\left(\theta_i,p_i^{(1)}\right)+\ve^2 Q_i^{(2)}\left(\theta_i,p_i^{(1)},p_i^{(2)}\right) \right.}\\
    &\quad\quad\quad\quad\quad\yp{\left.\left.+\ve^3 Q_i^{(3)}\left(\theta_i,p_i^{(1)},p_i^{(2)},p_i^{(3)}\right)+\cdots\right]\right\}}\\
    &\quad+ \ve\sum_{j=1}^N  a_{ij}\left\{\left[ I_i^{(0)}(\hat\theta_i+t) + \psi_i I_i^{(1)}(\hat\theta_i+t) + \psi_i^2 I_i^{(2)}(\hat\theta_i+t) +\cdots\right]\right.\\
    &\quad\quad\quad\quad\quad\quad  \cdot\left[ K_{ij}^{(0)}(\hat\theta_i+t,\hat\theta_j+t)\right.\\
    &\quad\quad\quad\quad\quad\quad\quad  + \ve K_{ij}^{(1)}(\hat\theta_i+t,\hat\theta_j+t,p_i^{(1)},p_j^{(1)}) \\
    &\quad\quad\quad\quad\quad\quad\quad  + \ve^2 K_{ij}^{(2)}(\hat\theta_i+t,\hat\theta_j+t,p_i^{(1)},p_i^{(2)},p_j^{(1)},p_j^{(2)}) \\
    &\quad\quad\quad\quad\quad\quad\quad  + \left.\left.\ve^3 K_{ij}^{(3)}(\hat\theta_i+t,\hat\theta_j+t,p_i^{(1)},p_i^{(2)},p_i^{(3)},p_j^{(1)},p_j^{(2)},p_j^{(3)}) + \cdots\right]\right\}.
\end{align*}
By writing $\ypb{\delta = \ve \delta/\ve \equiv \ve b}$ and re-collecting in powers of $\ve$ yields a hierarchy of scalar ODEs, which we show up to order $\ve^3$ below \eqref{eq:p_i_odes}. \yp{We suppress the explicit $\theta_i$-, $\theta_j$-dependence of $p_i^{(k)}$, $I^{(k)}$, and $K^{(k)}$}.
\begin{equation}\label{eq:p_i_odes}
    \begin{split}
        O(\ve)&:\quad \frac{dp_i^{(1)}}{dt} = \kappa_i p_i^{(1)} + \yp{\ypb{b} I_i^{(0)}\cdot Q_i^{(0)}} + \sum_{j=1}^N  a_{ij} I_i^{(0)} \cdot K_{ij}^{(0)},\\
        O(\ve^2)&:\quad\frac{dp_i^{(2)}}{dt} = \kappa_i p_i^{(2)} + \yp{\ypb{b} p_i^{(1)} I_i^{(1)}\cdot Q_i^{(0)} + \ypb{b}I_i^{(0)}\cdot Q_i^{(1)}} + \sum_{j=1}^N  a_{ij} \left(I_i^{(0)} \cdot K_{ij}^{(1)} + p_i^{(1)} I_i^{(1)}\cdot K_{ij}^{(0)}\right),\\
        O(\ve^3)&:\quad\frac{dp_i^{(3)}}{dt} = \kappa_i p_i^{(3)} +\yp{\ypb{b} p_i^{(2)}I_i^{(1)} + \ypb{b}\left(p_i^{(1)}\right)^2I_i^{(2)} + \ypb{b} p_i^{(1)}I_i^{(1)}\cdot Q_i^{(1)}+\ypb{b}I_i^{(0)}\cdot Q_i^{(2)}}\\
        &\quad\quad\quad\quad+ \sum_{j=1}^N  a_{ij} \left( I_i^{(0)} \cdot K_{ij}^{(2)} + p_i^{(1)}I_i^{(1)}\cdot K_{ij}^{(1)}+ p_i^{(2)}I_i^{(1)}\cdot K_{ij}^{(0)}   + \left(p_i^{(1)}\right)^2 I_i^{(2)}\cdot K_{ij}^{(0)}\right),\\
        &\vdots
    \end{split}
\end{equation}
Each \yp{ODE} is a first-order inhomogeneous differential equation with a forcing term that depends on lower-order solutions, and is thus solvable. To demonstrate our calculations concretely, we begin by explicitly solving the lowest-order term $p_i^{(1)}$. Throughout the calculations in the remainder of this section, we will often use $\theta_i$ in place of $\hat\theta_i + t$.

With all of its dependencies, the lowest-order equation is written
\begin{equation*}
    \frac{dp_i^{(1)}}{dt}(t) = \kappa_i p_i^{(1)}(t) + \ypb{b}I_i^{(0)}(\theta_i) \cdot Q_i^{(0)}(\theta_i) + \sum_{j_1=1}^N  a_{ij_1} I_i^{(0)}(\theta_i) \cdot K_{ij_1}^{(0)}(\theta_i,\theta_{j_1}).
\end{equation*}
Applying the integrating factor method, we arrive at the solution,
\begin{align*}
    p_i^{(1)}(t) &= \int_{t_0}^t e^{\kappa_i(t-s)}\left( \ypb{b} I_i^{(0)}(\hat\theta_i + s) \cdot Q_i^{(0)}(\hat\theta_i+s)+\sum_{j_1=1}^N  a_{ij_1}  I_i^{(0)}(\hat\theta_i+s) \cdot K_{ij_1}^{(0)}(\hat\theta_i+s,\hat\theta_{j_1}+s)\right) \dd s\\
    &\quad\quad+ e^{\kappa_i t}C
\end{align*}
where $C$ is a constant of integration. To discard transients, we ignore the constant of integration and take $t_0 \rightarrow -\infty$. Then, with the change of variables $s\to t-s$, the equation becomes,
\begin{equation}\label{eq:p1}
    \begin{split}
        p_i^{(1)}(t) = \ypb{b} q_i(\theta_i) + \sum_{j_1=1}^N a_{ij_1} g_{ij_1}(\theta_i,\theta_{j_1}),
    \end{split}
\end{equation}
where
\begin{align*}
    q_i(\xi) &= \ies I_i^{(0)}(\xi- s) \cdot Q_i^{(0)}(\xi-s) \, \dd s,\\
    g_{ij}(\xi_1,\xi_2) &=\ies I_i^{(0)}(\xi_1 -s) \cdot K_{ij}^{(0)}(\xi_1-s,\xi_2-s) \dd s,
\end{align*}
are functions that may be pre-computed. We use \eqref{eq:p1} to solve for $p_i^{(2)}$, yielding, with appropriate rearrangement,
\begin{equation}\label{eq:p2}
    p_i^{(2)}(t) = \underbrace{\ypb{b}\sum_{j_1} q_{ij_1}(\theta_i,\theta_{j_1})}_{\text{Heterogeneity}}+\underbrace{ \sum_{j_1,j_2} g_{ij_1j_2}(\theta_i,\theta_{j_1},\theta_{j_2})}_{\text{Coupling}},
\end{equation}
where each function $q_{ij_1}$ ($g_{ij_1j_2}$) is distinguishable from $q_i$ ($g_{ij_1}$)  by the number of indices. We find that the heterogeneous term contains two-body interactions and the coupling term contains three-body interactions. While our goal is not necessarily to enumerate every term explicitly, we show the forcing function terms for $p_i^{(2)}$ term by term \ypb{(and obtain more explicit equations for $q_{ij_1}$ and $g_{ij_1j_2}$ in \eqref{eq:pi2})}:
\begin{align*}
        \ypb{b} p_i^{(1)} I_i^{(1)} \cdot Q_i^{(0)} &= \ypb{b}^2 q_iI_i^{(1)} \cdot Q_i^{(0)} + \ypb{b}\sum_{j_1} a_{ij_1}g_{ij_1}I_i^{(1)} \cdot Q_i^{(0)},\\
        \ypb{b}I_i^{(0)} \cdot Q_i^{(1)} &= \ypb{b}I_i^{(0)}\cdot u_i^{(1)}[\ypb{b}q_i + \sum_{j_1}a_{ij_1} g_{ij_1}]\\
        &= \ypb{b}^2I_i^{(0)}\cdot u_i^{(1)}q_i + \ypb{b}\sum_{j_1}a_{ij_1} g_{ij_1}I_i^{(0)}\cdot u_i^{(1)},\\
        I_i^{(0)}\cdot K_{ij}^{(1)} &\equiv I_i^{(0)}\cdot [p_i^{(1)}v_{ij}^{(1,0)} + p_j^{(1)}v_{ij}^{(0,1)}]\\
        &=I_i^{(0)}\cdot [[\ypb{b}q_i + \sum_{j_1}a_{ij_1} g_{ij_1}]v_{ij}^{(1,0)} + [b_jq_j + \sum_{j_1}a_{jj_1} g_{jj_1}]v_{ij}^{(0,1)}]\\
        &= I_i^{(0)}\cdot [\ypb{b} q_iv_{ij}^{(1,0)} + b_jq_jv_{ij}^{(0,1)}] + I_i^{(0)}\cdot \sum_{j_1} [a_{ij_1}g_{ij_1}v_{ij}^{(1,0)} + a_{jj_1}g_{jj_1}v_{ij}^{(0,1)}],\\
        p_i^{(1)}I_i^{(1)}\cdot K_{ij}^{(0)} &= [\ypb{b}q_i + \sum_{j_1}a_{ij_1} g_{ij_1}] I_i^{(1)}\cdot K_{ij}^{(0)}\\
        &= \ypb{b}q_iI_i^{(1)}\cdot K_{ij}^{(0)} + \sum_{j_1}a_{ij_1} g_{ij_1}I_i^{(1)}\cdot K_{ij}^{(0)}.
\end{align*}
Thus, we may write \eqref{eq:p2} more explicitly:
\begin{align}\label{eq:pi2}
        p_i^{(2)}(t) = b^2 q_{ij_0}(\theta_i) + b \sum_{j_1} q_{ij_1}(\theta_i,\theta_{j_1})+\sum_{j_1}\sum_{j_2}g_{ij_1j_2}(\theta_i,\theta_{j_1},\theta_{j_2}) + \hat g_{ij_1j_2}(\theta_i,\theta_{j_1},\theta_{j_2}) ,
\end{align}
where
\begin{align*}
    q_{ij_0}(\xi) &= \ies \left[ q_i(\xi-s)I_i^{(1)}(\xi-s) \cdot Q_i^{(0)}(\xi-s)\right.\\
    &\quad\quad\quad\quad\quad\quad\left.+  q_i(\xi-s) I_i^{(0)}(\xi-s)\cdot u_i^{(1)}(\xi-s) \right]\dd s,\\
    q_{ij_1}(\xi_1,\xi_2) &= \ies \left[q_i(\xi_1-s) I_i^{(0)}(\xi_1-s)\cdot v_{ij_1}^{(1,0)}(\xi_1-s,\xi_2-s)\right. \\
    &\quad\quad\quad\quad\quad\quad+ q_{j_1}(\xi_2-s) I_i^{(0)}(\xi_1-s)\cdot v_{ij_1}^{(0,1)}(\xi_1-s,\xi_2-s)\\
    &\quad\quad\quad\quad\quad\quad+ q_i(\xi_1-s)I_i^{(1)}(\xi_1-s)\cdot K_{ij_1}^{(0)}(\xi_1-s,\xi_2-s)\\
    &\quad\quad\quad\quad\quad\quad+ g_{ij_1}(\xi_1-s,\xi_2-s)I_i^{(1)}(\xi_1-s) \cdot Q_i^{(0)}(\xi_1-s)\\
    &\quad\quad\quad\quad\quad\quad\left.+ g_{ij_1}(\xi_1-s,\xi_2-s)I_i^{(0)}(\xi_1-s) \cdot u_i^{(1)}(\xi_1-s)\right]\dd s,\\
    g_{ij_1j_2}(\xi_1,\xi_2,\xi_3) &= \ies \left[ g_{ij_2}(\xi_1-s,\xi_3-s)I_i^{(0)}(\xi_1-s)\cdot v_{ij_1}^{(1,0)}(\xi_1-s,\xi_2-s)\right.\\
    &\quad\quad\quad\quad\quad\quad+\left. g_{ij_2}(\xi_1-s,\xi_3-s)I_i^{(1)}(\xi_1-s)\cdot K_{ij_1}^{(0)}(\xi_1-s,\xi_2-s) \right]\dd s,\\
    \hat g_{ij_1j_2}(\xi_1,\xi_2,\xi_3) &= \ies   g_{j_1j_2}(\xi_2-s,\xi_3-s)I_i^{(0)}(\xi_1-s)\cdot v_{ij_1}^{(0,1)}(\xi_1-s,\xi_2-s) \dd s.
\end{align*}
\ypb{Even more explicit terms that include pairwise coupling terms $a_{ij}$ are shown in \cite{nicks2024insights} Appendix C.}

Note that $b$ increases in power for each new term, thus we require $b\leq 1$ for arbitrarily high-order expansions, which is equivalent to the condition that the overall strength of heterogeneity must be less than or equal to the overall coupling strength ($\delta \leq \ve$). This condition is not restrictive, however, because we tend not to take arbitrarily high expansions, and coupling may be weakened relative to heterogeneity by adjusting the coupling constants $a_{ij}$.

We continue these calculations for each order $k$ to obtain solutions of the same form as \eqref{eq:p2}:
\begin{align*}
    p_i^{(k)}(t) = &\underbrace{ b\sum_{\ell=1}^k \sum_{j_1,\ldots,j_{\ell-1}}q_{ij_1\cdots j_{\ell-1}}(\theta_i,\theta_{j_1},\ldots,\theta_{j_{\ell-1}})}_{\text{Heterogeneity}}+ \underbrace{\sum_{\ell=1}^k\sum_{j_1,\ldots,j_\ell} g_{ij_1\cdots j_\ell}(\theta_i,\theta_{j_1},\ldots,\theta_{j_\ell})}_{\text{Coupling}}.
\end{align*}
The maximum number of indices $j_1,\ldots,j_k$ is $N-1$, because we use one index for oscillator $i$ and there are $N$ oscillators. Naturally, higher-order terms for $k\geq N$ do not introduce additional higher-order interactions, but may improve the accuracy of the expansion.

Now that we have expressed each $p_i^{(k)}$ in terms of phase, effectively eliminating the isostable equation, we turn to deriving a phase difference equation.

\subsection{Phase Difference Equation}\label{sec:phase_diff}

We expand the phase equation \eqref{eq:th_moving} in $\ve$ \yp{(where again, for brevity, not all $\psi_i$ are shown expanded in $\ve$)}:
\begin{align*}
    \dot{\hat\theta}_i &= \yp{\ypb{\delta} \mathcal{Z}_i(\hat\theta_i+t,\psi_i) \cdot Q_i(\hat\theta_i+t,\psi_i)} + \ve \sum_{j=1}^N a_{ij} \mathcal{Z}_i (\hat\theta_i + t,\psi_i) \cdot G_{ij}(\hat\theta_i+t,\hat\theta_j+t) \\
    &= \yp{\ypb{\delta} \left\{\left[ Z_i^{(0)}(\hat\theta_i+t) + \psi_i Z_i^{(1)}(\hat\theta_i+t) + \psi_i^2 Z_i^{(2)}(\hat\theta_i+t) +\cdots\right]\right.}\\
    &\quad\quad\quad\quad\yp{\cdot\left[ Q_i^{(0)}(\theta_i)+\ve Q_i^{(1)}\left(\theta_i,p_i^{(1)}\right)+\ve^2 Q_i^{(2)}\left(\theta_i,p_i^{(1)},p_i^{(2)}\right) \right.}\\
    &\quad\quad\quad\quad\quad\yp{\left.\left.+\ve^3 Q_i^{(3)}\left(\theta_i,p_i^{(1)},p_i^{(2)},p_i^{(3)}\right)+\cdots\right]\right\}}\\
    &\quad+\ve \sum_{j=1}^N a_{ij} \left[Z_i^{(0)}+\psi_i Z_i^{(1)} + \psi_i^2 Z_i^{(2)}+\cdots\right]\cdot \left[K_{ij}^{(0)} + \ve K_{ij}^{(1)} + \ve^2 K_{ij}^{(2)}+\cdots\right].
\end{align*}
Substituting the expansion for $\psi_i$ and collecting in powers of $\ve$ yields a virtually identical right-hand side as \eqref{eq:p_i_odes} with $Z$ in place of $I$:
\begin{align}\label{eq:thetahat}
        \dot{\hat\theta}_i &=\ve\left[ \yp{b Z_i^{(0)}\cdot Q_i^{(0)}} + \sum_{j=1}^N a_{ij} Z_i^{(0)}\cdot K_{ij}^{(0)} \right]\\
        &\quad +\ve^2 \left[\yp{b p_i^{(1)} Z_i^{(1)}\cdot Q_i^{(0)} + b Z_i^{(0)}\cdot Q_i^{(1)}} + \sum_{j=1}^N a_{ij}\left( Z_i^{(0)}\cdot K_{ij}^{(1)} + p_i^{(1)} Z_i^{(1)}\cdot K_{ij}^{(0)}\right)\right]\nonumber \\
        &\quad +\ve^3\left[\vphantom{\sum_{j=1}^N}\yp{b p_i^{(2)}I_i^{(1)} + b\left(p_i^{(1)}\right)^2I_i^{(2)} + b p_i^{(1)}I_i^{(1)}\cdot Q_i^{(1)}+bI_i^{(0)}\cdot Q_i^{(2)}}\right.\nonumber\\
        &\quad\quad\quad + \left.\sum_{j=1}^N a_{ij}\left( Z_i^{(0)}\cdot K_{ij}^{(2)} + p_i^{(1)} Z_i^{(1)}\cdot K_{ij}^{(1)}  +  p_i^{(2)}Z_i^{(1)} \cdot K_{ij}^{(0)}   + \left( p_i^{(1)}\right)^2 Z_i^{(2)} \cdot K_{ij}^{(0)} \right)\right]\nonumber\\
        &\,\,\,\vdots.\nonumber
\end{align}
This differential equation is a system of nonautonomous ODEs for the phase dynamics of each oscillator.

We now use averaging theory to transform this non-autonomous system \eqref{eq:thetahat} into an autonomous system. Again, for concreteness, we examine the first few terms of the right-hand side. The averaged order $O(\ve)$ dynamics satisfy
\begin{equation*}
    b \tilde q_i +  \sum_{j_1=1}^N a_{ij_1} \mathcal{H}_{ij_1}^{(1)}(\theta_{j_1}-\theta_i).
\end{equation*}
where
\begin{align*}
    \tilde q_i &= \frac{1}{T} \int_0^T Z_i^{(0)}(s)\cdot Q_i^{(0)}(s) \dd s,\\
    \mathcal{H}_{ij_1}(\xi) &= \frac{1}{T}\int_0^T Z_i^{(0)}(s)\cdot K_{ij_1}^{(0)}(s,\xi+s)\dd s.
\end{align*}
The $\mathcal{H}_{ij_1}^{(1)}$ term is identical to the classic weak coupling theory for $N$ oscillators (see, e.g., \cite{ermentrout2010} Chapter 8 Equation (8.38)). The next order $O(\ve^2)$ terms are,
\begin{align*}
    b p_i^{(1)} Z_i^{(1)} \cdot Q_i^{(0)} &= b^2 \tilde q_i Z_i^{(1)} \cdot Q_i^{(0)} + b\sum_{j_1} a_{ij_1}g_{ij_1}Z_i^{(1)} \cdot Q_i^{(0)},\\
    bZ_i^{(0)} \cdot Q_i^{(1)} &= bZ_i^{(0)}\cdot u_i^{(0)}[b\tilde q_i + \sum_{j_1}a_{ij_1} g_{ij_1}]\\
    &= b^2Z_i^{(0)}\cdot u_i^{(0)}\tilde q_i + b\sum_{j_1}a_{ij_1} g_{ij_1}Z_i^{(0)}\cdot u_i^{(0)},\\
    Z_i^{(0)}\cdot K_{ij}^{(1)} &\equiv Z_i^{(0)}\cdot [p_i^{(1)}v_{ij}^{(1,0)} + p_j^{(1)}v_{ij}^{(0,1)}]\\
    &=Z_i^{(0)}\cdot [[b\tilde q_i + \sum_{j_1}a_{ij_1} g_{ij_1}]v_{ij}^{(1,0)} + [bq_j + \sum_{j_1}a_{jj_1} g_{jj_1}]v_{ij}^{(0,1)}]\\
    &= bZ_i^{(0)}\cdot [\tilde q_i v_{ij}^{(1,0)} + q_jv_{ij}^{(0,1)}] + Z_i^{(0)}\cdot \sum_{j_1} a_{ij_1}g_{ij_1}v_{ij}^{(1,0)} + a_{jj_1} g_{jj_1}v_{ij}^{(0,1)},\\
    p_i^{(1)}Z_i^{(1)}\cdot K_{ij}^{(0)} &= [b\tilde q_i + \sum_{j_1}a_{ij_1} g_{ij_1}] Z_i^{(1)}\cdot K_{ij}^{(0)}\\
    &= b\tilde q_i Z_i^{(1)}\cdot K_{ij}^{(0)} + \sum_{j_1}a_{ij_1} g_{ij_1}Z_i^{(1)}\cdot K_{ij}^{(0)}.
\end{align*}
Thus, the second-order term is given by
\begin{align*}
    b\sum_{j_1} \tilde q_{ij_1}(\hat\theta_{j_1}-\hat\theta_i) + \sum_{j_1,j_2} \mathcal{H}_{ij_1j_2}(\hat\theta_{j_1}-\hat\theta_i,\hat\theta_{j_2}-\hat\theta_i),
\end{align*}
where
\begin{align*}
    b\sum_{j_1} \tilde q_{ij_1}(\xi) = &\frac{b^2}{T}\int_0^T  \tilde q_i(s)Z_i^{(1)}(s) \cdot Q_i^{(0)}(s) \dd s+ \frac{b^2}{T}\int_0^T  \tilde q_i(s)u_i^{(1)}(s) \cdot Q_i^{(0)}(s)\dd s\\
    &+\sum_{j_1}\frac{b}{T}\int_0^T \tilde q_i(s) Z_i^{(0)}(s)\cdot v_{ij_1}^{(1,0)}(s,\xi+s) \dd s\\
    &+\sum_{j_1}\frac{b}{T}\int_0^T \tilde q_{j_1}(\xi+s) Z_i^{(0)}(s)\cdot v_{ij_1}^{(0,1)}(s,\xi+s) \dd s\\
    &+\sum_{j_1} \frac{b}{T}\int_0^T \tilde q_i(s)Z_i^{(1)}(s)\cdot K_{ij_1}^{(0)}(s,\xi+s) \dd s\\
    &+\sum_{j_1} \frac{ba_{ij_1}}{T}\int_0^T Z_i^{(1)}(s) \cdot Q_i^{(0)}(s)g_{ij_1}(s,\xi+s)\dd s\\
    &+\sum_{j_1}\frac{ba_{ij_1} }{T}\int_0^T  Z_i^{(1)}(s) \cdot Q_i^{(0)}(s)g_{ij_1}(s,\xi+s)\dd s,\\
    \sum_{j_1,j_2} \mathcal{H}_{ij_1j_2}(\xi_1,\xi_2) =&\sum_{j_1,j_2}  \frac{a_{ij_2}}{T}\int_0^T Z_i^{(0)}(s)\cdot v_{ij_1}^{(1,0)}(s,\xi+s)  g_{ij_2}(s,\xi+s)\dd s\\
    &+\sum_{j_1,j_2}  \frac{a_{j_1j_2}}{T}\int_0^T Z_i^{(0)}(s)\cdot v_{ij_1}^{(1,0)}(s,\xi_1+s)  g_{j_1j_2}(\xi_1+s,\xi_2+s)\dd s\\
    &+\sum_{j_1,j_2}  \frac{a_{ij_2}}{T}\int_0^T Z_i^{(0)}(s)\cdot K_{ij_1}^{(0)}(s,\xi_1+s)  g_{ij_2}(s,\xi_2+s)\dd s.
\end{align*}
Three-body interactions are apparent, as are the existence of interactions between coupling and heterogeneity. These terms, with $b=0$, are the second-order interaction terms from \cite{wilson2019phase}. A careful examination of these explicit terms may be of interest in future work, but we collapse them into single functions and continue these calculations with the aid of a symbolic package up to some chosen order $k$:
\begin{equation*}
    \dot{\hat\theta}_i =  \underbrace{ b\sum_{\ell=1}^k \ve^\ell \sum_{j_1,\ldots,j_{\ell-1}}\tilde q_{ij_1\cdots j_{\ell-1}}^{(k)}(\hat\theta_i,\hat\theta_{j_1},\ldots,\hat\theta_{j_{\ell-1}})}_{\text{Heterogeneity}}+ \underbrace{\sum_{\ell=1}^k \ve^\ell \sum_{j_1,\ldots,j_\ell} \mathcal{H}_{ij_1\cdots j_\ell}^{(k)}(\hat\theta_i,\hat\theta_{j_1},\ldots,\hat\theta_{j_\ell})}_{\text{Coupling}}.
\end{equation*}
Defining $\phi_i = \hat\theta_i - \hat\theta_1$, we arrive at the phase difference equation,
\begin{equation}
    \begin{split}
        \dot{\phi}_i &=b\sum_{\ell=1}^k \ve^\ell \sum_{j_1,\ldots,j_{\ell-1}} \left[ \tilde{q}_{i,j_1,\ldots,j_{\ell-1}}^{(\ell)}(\phi_{j_1}-\phi_i,\ldots,\phi_{j_{\ell-1}}-\phi_i) - \tilde{q}_{1,j_1,\ldots,j_{\ell-1}}^{(\ell)}(\phi_{j_1},\ldots,\phi_{j_{\ell-1}})\right],\\
        &+\sum_{\ell=1}^k \ve^\ell  \sum_{j_1,\ldots,j_\ell} \left[ \mathcal{H}_{i,j_1,\ldots,j_\ell}^{(\ell)}(\phi_{j_1}-\phi_i,\ldots,\phi_{j_\ell}-\phi_i) - \mathcal{H}_{1,j_1,\ldots,j_\ell}^{(\ell)}(\phi_{j_1},\ldots,\phi_{j_\ell})\right],
    \end{split}
\end{equation}
for $i=2,\ldots,N$. This equation is a generalized version of the two-body interactions in \cite{park2021high} and a generalization beyond the second-order coupling terms in \cite{wilson2019phase}. Since we will not examine particular terms in the inner summation, we rewrite the right-hand side as a single $\mathcal{H}$ function for each oscillator $i$ and order $\ell$ for notational convenience:
\begin{equation}\label{eq:phi}
    \dot{\phi}_i = \sum_{\ell=1}^k\ve^\ell  \left[b\tilde{q}_i^{(\ell)}(\phi_{2},\ldots,\phi_{N}) + \mathcal{H}_{i}^{(\ell)}(\phi_{2},\ldots,\phi_{N}) \right],\quad i=2,\ldots,N.
\end{equation}

\paragraph{Remarks}
\begin{enumerate}
    \item At order $k$ and above, heterogeneity contributes $(k-1)$-body interactions (heterogeneity contributes a constant term in the lowest-order term).
    \item The choice of additive, linear heterogeneity in the original system \eqref{eq:odes}, yields additive, linear heterogeneity in the reduced dynamics \eqref{eq:phi}. This choice is not strictly necessary.  While we do not require oscillators to have identical or even similar vector fields, consider for the sake of example that we have a system of identical neural oscillators \ypb{where oscillator $i$ has one heterogeneous parameter $c_i$ in the dynamics of some gating variable $X$,
    \begin{equation*}
        X_\infty(V_i;c_i) = 1/(1+\exp((V_i+c_i))),
    \end{equation*}
    while all other oscillators $j\neq i$ have $c_i=0$. Such a system can be transformed into the current framework by taking a Taylor expansion from the parameter value(s) $c^*=0$ at which all oscillator periods are the same. Then,
    \begin{equation*}
        X_\infty(V_i;c_i) = X_\infty(V_i;c^*) + (c_i-c^*) X_1 + (c_i-c^*)^2 X_2 + \cdots,
    \end{equation*}
    where $X_i$ are the higher-order terms in the Taylor expansion. Taking $\delta = c_i$, higher-order $\delta$ terms may be included in the formulation.}
    \item \ypb{While we simplified our assumptions to include only one heterogeneous parameter, a generalization to multiple heterogeneous parameters is straightforward. Using the same calculations as above, if $c^*$ is an arbitrary heterogeneous term where all oscillators have equal period, then we can compute the same Taylor expansion as above for each oscillator and obtain the parameter $\delta_i=|c_i-c^*|$ for each $i$. Note that there is no restriction on the number of heterogeneous parameters per oscillator, so long as there exist parameter values where all oscillator periods are equal.}
    \item Heterogeneity and asymmetry may induce interesting dynamics that will be considered in future work, but we disregard heterogeneous effects for the remainder of the paper to focus on the primary goal of confirming that the proposed $N$-body phase reduction is valid in the simple case of identical oscillators with homogeneous coupling.
    \item The phase difference equation may also be derived by using the averaged isostable expansion terms, $\bar p_i^{(k)}$, which, like $\hat\theta_i$, tends to have a slowly varying mean, even when $\ve$ is not arbitrarily small. The average term $\bar p_i^{(k)}$ is useful because it only depends on the relatively slow phases $\hat\theta_i,\hat\theta_j$, so it may be moved out of time integrals, greatly simplifying calculations. We perform this type of averaging in the thalamic model in Section \ref{sec:thal}.
\end{enumerate}

\section{Complex Ginzburg-Landau Model}\label{sec:cgl}
We now apply the proposed method to a \ypb{set of three globally coupled complex Ginzburg-Landau (CGL) models, where the coupling is diffusive and homogeneous}. The ODE form of this model has been studied extensively \cite{wilson2019phase,park2021high}, making it an ideal preliminary test for our results. Let $X_i = (x_i,y_i)^\tr$ and $N=3$. The network is given by
\begin{equation*}
    \dot X_i = F(X_i) + \sum_{j=1}^3 \yp{a_{ij}} G(X_i,X_j),
\end{equation*}
where
\begin{equation*}
    F(X_i) = \left(\begin{matrix}
        \sigma x_i(1-R_i) - y_i(1+\rho(R_i-1))\\
        \sigma y_i(1-R_i) + x_i(1+\rho(R_i-1))
    \end{matrix}\right), \quad
    G(X_i,X_j) = \left(\begin{matrix}
        (x_j-x_i) - d(y_j - y_i)\\
        (y_j-y_i) + d (x_j - x_i)
    \end{matrix}\right),
\end{equation*}
and $R_i=x_i^2+y_i^2$. We assume all-to-all coupling without self coupling, so that pairwise terms $a_{ij}$ are given by,
\begin{equation*}
    a_{ij}=\begin{cases}
        1/N &\text{if } i\neq j\\
        0 & \text{else}
    \end{cases}.
\end{equation*}
\ypb{It is straightforward to verify that the Floquet exponent for this system is given by $\kappa = -2\sigma$, where the Floquet multiplier $\kappa$ no longer has a subscript because all oscillators are identical in this example.}

\subsection{Explicit Calculations}\label{sec:cgl_explicit}

The $g$, $Z$, and $I$ functions of the CGL model each have only one (the first) nontrivial mode. To make explicit calculations, we write these functions in complex Fourier form,
\begin{align*}
    g_i^{(k)}(\theta_i) &= \left[\begin{array}{l}
        a_{g}^{(k)} \exp(\ypb{\mi}\theta_i) + \text{c.c.}\\
        b_{g}^{(k)} \exp(\ypb{\mi}\theta_i) + \text{c.c.}
    \end{array}\right],\\
    Z_i^{(k)}(\theta_i) &= \left[\begin{array}{l}
        a_{Z}^{(k)} \exp(\ypb{\mi}\theta_i) + \text{c.c.}\\
        b_{Z}^{(k)} \exp(\ypb{\mi}\theta_i) + \text{c.c.}
    \end{array}\right],\\
    I_i^{(k)}(\theta_i) &= \left[\begin{array}{l}
        a_{I}^{(k)} \exp(\ypb{\mi}\theta_i) + \text{c.c.}\\
        b_{I}^{(k)} \exp(\ypb{\mi}\theta_i) + \text{c.c.}
    \end{array}\right],
\end{align*}
where $a_X^{(k)},b_X^{(k)}\in\mathbb{C}$, $X\in\{g,Z,I\}$, and ``c.c.''~stands for ``complex conjugate.'' \ypb{We use the upright $\mi$ for imaginary numbers and keep the italicized $i$ for indices}. Note that due to symmetry, $b_X^{(k)} = -\ypb{\mi} a_X^{(k)}$ for all $k$. The limit cycle coefficients are simply $a_L=1/2$ and $b_L=-\ypb{\mi}/2$. 

By assumption, $p_i^{(0)}=0$, so we calculate the next order term,
\ypb{\begin{align*}
    p_{i}^{(1)}(\hat\theta_1,\hat\theta_2,\hat\theta_3) = \frac{1}{3\kappa}\sum_{j=1}^3 \left[ c_1 (1-e^{\ypb{\mi(\hat\theta_i-\hat\theta_j)}}) + \text{c.c.}\right] ,
\end{align*}
where $c_1 = 2 a_I^{(0)} \bar{a}_L (1-\mi d) \in\mathbb{C}$}. Continuing the calculation,
\ypb{\begin{align*}
    &p_{i}^{(2)}(\hat\theta_1,\hat\theta_2,\hat\theta_3) =  \frac{4}{9\kappa^2}\sum_{j_1=1}^3\sum_{j_2=1}^3\left[c_2 e^{\mi({\hat\theta_{i}-\hat\theta_{j_1}})} (\Re(c_1 e^{{\mi}({\hat\theta_{j_1}-\hat\theta_{j_2}})}) -\Re(c_1))+\text{c.c.}\right] \\
    &+\frac{4}{9\kappa^2}\sum_{j_1=1}^3 \sum_{j_2=1}^3 \left[   d_1 e^{\ypb{\mi}(\hat\theta_i-\hat\theta_{j_1})} + d_1 e^{\ypb{\mi}(\hat\theta_i-\hat\theta_{j_2})} + \Re(d_2)e^{\ypb{\mi}(\hat\theta_{j_1}-\hat\theta_{j_2})} + d_2e^{\ypb{\mi}( 2\hat\theta_i -\hat\theta_{j_1}-\hat\theta_{j_2})}+d_3 + \text{c.c} \right]
\end{align*}}
where \ypb{$d_1=-( \Re(c_1c_2+ d_2)+d_2)$, $d_2=c_1c_3$, $d_3=2c_1 \Re(c_2+c_3)$, $c_2=a_I^{(0)}\bar a_g^{(1)}(1-\mi d)$, $c_3=a_I^{(1)} \bar a_L(1-\mi d)$.  The function $\Re$ simply returns the real part of its input.}

\paragraph{Remarks}
\begin{itemize}
    \item The nontrivial Floquet exponent, $\kappa$, appears in the denominator of both $p_i^{(1)}$ and $p_i^{(2)}$. Thus, a smaller nontrivial Floquet exponent ($|\kappa| \ll 1$), which occurs for weakly attracting limit cycles, makes the isostable coordinate contribute nontrivially to the phase dynamics. While we generically expect $\kappa$ to appear implicitly in the denominator, these calculations make its appearance explicit. Indeed, the inclusion of isostable coordinates for $|\kappa|$ small is the key difference between our work and purely phase-based methods.
    \item \ypb{The coefficient for each $p_i^{(k)}$ is of the form $1/(N\kappa)^k$ and thus may diverge for large $k$. Questions of convergence will be explored in future work, but we note that this issue does not appear to significantly affect the results for the relatively lower-order truncation we consider.}
    \item Higher-order terms not only contain higher-order interactions, but higher-order Fourier modes, \ypb{e.g., $e^{\ypb{\mi}( 2\hat\theta_i -\hat\theta_{j_1}-\hat\theta_{j_2})}$.}
    \item \ypb{In this example, each function $p_i^{(1)}$ and  $p_i^{(2)}$ only depends on phase differences, so that $p_i^{(k)}(\hat\theta_1+t,\hat\theta_2+t,\hat\theta_3+t) = p_i^{(k)}(\hat\theta_1,\hat\theta_2,\hat\theta_3)$. This pure phase dependence holds so long as $b_X^{(k)} = -\ypb{\mi} a_X^{(k)}$ and the oscillators are identical -- we exploit this property for the numerical results in Section \ref{sec:cgl_num} (this lack of time dependence is a lucky coincidence. In general, even with identical oscillators, we expect to see additional terms that do not purely depend on phase differences). }
\end{itemize}

With the solutions $p_i^{(1)}$ and $p_i^{(2)}$ in hand, we turn to the calculation of $\mathcal{H}$ functions.
\begin{align*}
    \mathcal{H}_i^{(1)} = \ypb{\frac{1}{3}}\sum_{j=1}^3  \left[ \hat{c}_1  (e^{\ypb{\mi(\hat\theta_i-\hat\theta_j)}} -1)+ \text{c.c.}\right],
\end{align*}
where $\hat{c}_1=2 \ypb{a_Z^{(0)}\bar{a}_L}(1-\ypb{\mi}d)\in\mathbb{C}$.
\begin{align*}
    &\mathcal{H}_i^{(2)} =  \ypb{\frac{2}{9\kappa}\sum_{j_1=1}^3\sum_{j_2=1}^3   \left[\hat c_2(\Re(\hat c_1 e^{{\mi}({\hat\theta_{j_1}-\hat\theta_{j_2}})}) -\Re(\hat c_1)) + \text{c.c.}\right]} \\
    &+\ypb{\frac{1}{9\kappa}\sum_{j_1=1}^3 \sum_{j_2=1}^3 \left[  \hat{d}_1e^{\ypb{\mi}(\hat\theta_{j_1}-\hat\theta_i )} + \hat{d}_1 e^{\ypb{\mi}( \hat\theta_{j_2}-\hat\theta_i)} + \Re{(\hat{d}_2)}e^{\ypb{\mi}( \hat\theta_{j_2}-\hat\theta_{j_1})} + \hat d_2 e^{\ypb{\mi}( -2\hat\theta_i +\hat\theta_{j_2}+\hat\theta_{j_1})}+\hat{d}_4 +\text{c.c.}\right],}
\end{align*}
where \ypb{$\hat{d}_1=( \Re(\hat{c}_1\hat c_2+ \hat d_2)+\hat d_2)$, $\hat d_2=-\hat c_1 \hat c_3$, $\hat d_3=-2\hat c_1 \Re(\hat c_2+ \hat c_3)$, $\hat c_2 = a_Z^{(0)}\bar{a}_g^{(1)} (1-\ypb{\mi}d)$, $\hat{c}_3=a_Z^{(1)}\bar{a}_L(1-\ypb{\mi}d)$}. While it is possible to calculate $\mathcal{H}_i^{(3)}$ explicitly, the terms are far too numerous to show here. We instead proceed numerically.

\subsection{Numerical Results}\label{sec:cgl_num}
Numerically calculating the $\mathcal{H}$ functions \eqref{eq:phi} for the CGL model is straightforward because the Nyquist frequency of the underling function is especially low and requires only a dozen Fourier terms at $O(\ve^3)$. Using a Fourier truncation in this scenario greatly reduces the time complexity and memory requirements for the averaging calculation behind the $\mathcal{H}$ functions (see Appendix \ref{a:fourier} for additional details).

\begin{figure}[ht!]
    \includegraphics[width=\textwidth]{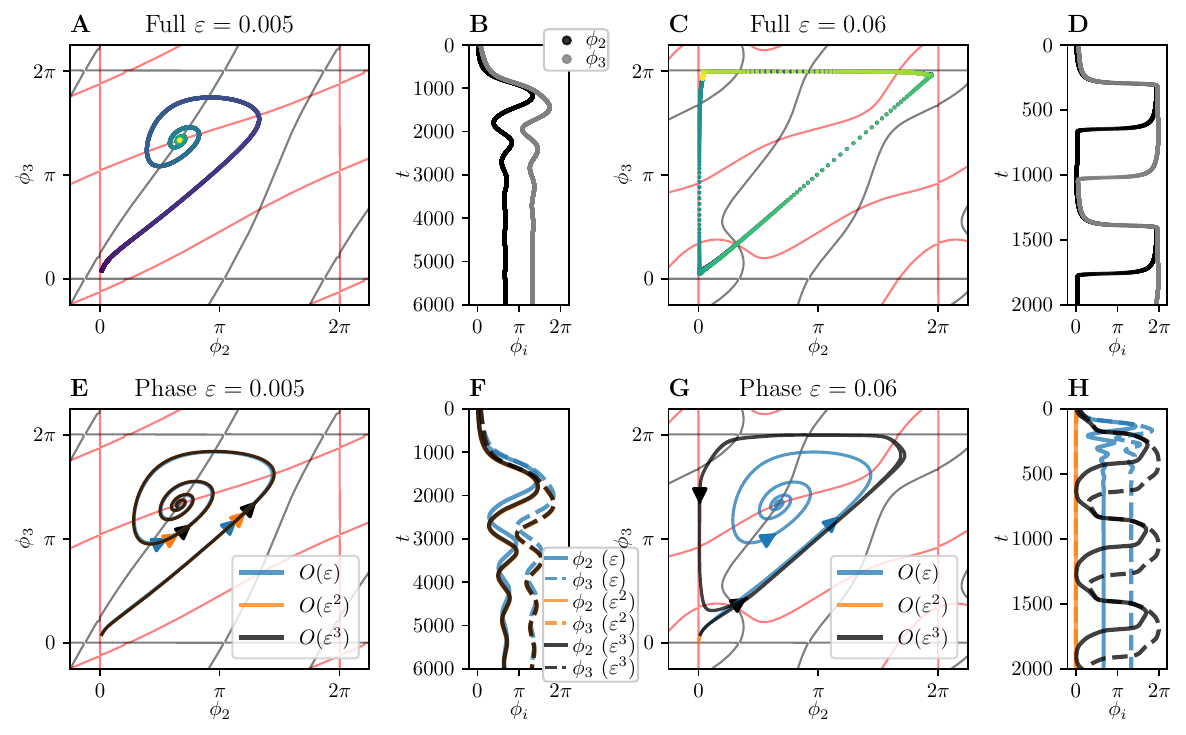}
    \caption{Comparison of the full (top row) and reduced (bottom row) CGL models. All panels show the corresponding nullclines of the $O(\ve^3)$ reduced model. A: Phase difference estimate of the full model dynamics at $\ve=0.005$. Lighter shades indicate later times. E: The corresponding reduced models ($O(\ve)$ blue, $O(\ve^2)$ orange, and $O(\ve^3)$ black). Arrows indicate movement in forward time. Note that all phase models coincide. B, F: Corresponding plots over time of the full model and reduced model, respectively. C, D, full and reduced model dynamics, respectively for $\ve=0.06$. D, H: corresponding plots over time of the full and reduced models, respectively. Parameters: $d=0.9$, $\sigma = 0.1$, $\rho = 0.15$. \ypb{We show every hundredth time point of the full model phase estimation to reduce lag when rendering vector graphics in this document}.}\label{fig:cgl_examples}
\end{figure}

We show comparisons between the full and reduced versions of the CGL model in Figure \ref{fig:cgl_examples}. The top row shows phase estimates of the full model for $\ve=0.005$ (A) and $\ve=0.06$ (C), where later shades correspond to later times \yp{(see Appendix \ref{a:phase_estimate} for more details on the phase estimation of the full model)}. The bottom row shows the $O(\ve)$ (blue), $O(\ve^2)$ (orange), and $O(\ve^3)$ (black) phase models exhibiting qualitatively similar dynamics at $\ve=0.005$ (E) and $\ve=0.06$ (G), respectively. Corresponding time traces are shown to the right of each portrait, e.g., panel B corresponds to A, and F corresponds to E.

At $\ve=0.005$, the full model tends towards an asymptotically stable splay state when initialized near synchrony with phases $(\phi_2,\phi_3)=(0.05,0.25)$ (A,B, where $\phi_i \in [0,2\pi)$). With the same initial values, all $O(\ve)$ (blue), $O(\ve^2)$ (orange), and $O(\ve^3)$ (black) phase models coincide with each other and with the full model as expected (E,F). For greater values of $\ve$, the full model splay state loses stability and the phase differences converge towards a limit cycle attractor (C,D). Only the $O(\ve^3)$ (black) phase model exhibits a similar limit cycle oscillation, while the $O(\ve)$ (blue) phase model dynamics does not change, and the $O(\ve^2)$ (orange) phase model simply converges to synchrony due to changes in the underlying basin of attraction (G,H).

\begin{figure}[ht!]
    \includegraphics[width=\textwidth]{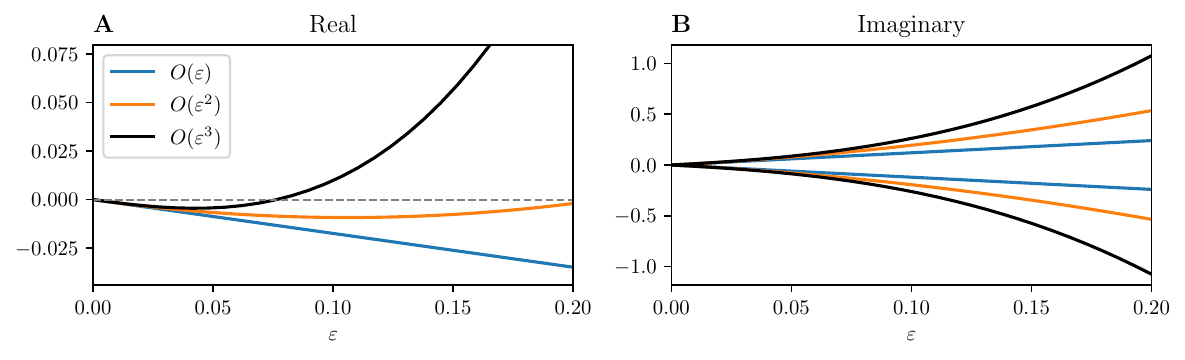}
    \caption{Real (A) and imaginary (B) parts of the eigenvalues of the Jacobian matrix evaluated at the splay state in the reduced CGL models. Blue: $O(\ve)$, orange: $O(\ve^2)$, black: $O(\ve^3)$. \yp{The dashed line is provided for reference to highlight the loss of stability as $\ve$ increases in higher order accuracy models.} Parameters are identical to those used in Figure \ref{fig:cgl_examples}.}\label{fig:cgl_eigs}
\end{figure}

Note that even though both $O(\ve^2)$ and $O(\ve^3)$ terms contain 3-body interactions, only the $O(\ve^3)$ phase model reproduces limit cycle behavior. This example demonstrates that $N$-body interactions alone are not always sufficient to capture the dynamics of the original model. Additional correction terms may be necessary.

The stability of the splay state is straightforward to calculate using the reduced model because only the eigenvalues of the Jacobian matrix evaluated at the splay state $(\phi_2,\phi_3) = (2\pi/3,4\pi/3)$ need to be known. By using the Fourier expansion of the $\mathcal{H}$ functions, only derivatives of sinusoids are required to compute the Jacobian, and this derivative can be taken rapidly without the need for estimates such as finite differences. The result of this analysis is shown in \yp{Figure} \ref{fig:cgl_eigs}. The left and right panels show the real and imaginary parts of the eigenvalues, respectively. While $O(\ve^2)$ (orange) does eventually lose stability, it occurs at a 4-fold greater coupling strength than the full or $O(\ve^3)$ models.


\section{Thalamic Model}\label{sec:thal}

We now apply the method to a set of $N=3$ synaptically coupled conductance-based thalamic neuron models from \cite{rubi04}. These results extend our previous work where we applied a strongly coupled phase reduction method for $N=2$ thalamic models \cite{park2021high}.

\yp{\paragraph{Remark} To simplify calculating the $p_i^{(k)}$ terms for this model, we compute the averaged dynamics $\bar p_i^{(k)}$. Then the integral calculations for each each $p_i^{(k)}$ become trivial, but the resulting solution nevertheless closely follows its original trajectory.}

\begin{table}
    \caption {Thalamic model parameter values.} \label{tab:thal}
    \begin{center}
        \begin{tabular}{l|l}
            Parameter & Value\\
            \hline
            $C$& \SI{1}{\micro F/cm^2}\\
            $E_k$& \SI{-90}{mV}\\
            $E_\text{Na}$& \SI{50}{mV}\\
            $E_\text{t}$& \SI{0}{mV}\\
            $E_\text{L}$& \SI{-70}{mV}\\
            $E_\text{syn}$& \SI{0}{mV} (Figure \ref{fig:thal_examples1}) or \SI{-100}{mV} (Figure \ref{fig:thal_examples2})\\
            $g_\text{L}$& \SI{0.05}{mS/cm^2}\\
            $g_\text{K}$& \SI{5}{mS/cm^2}\\
            $g_\text{Na}$& \SI{3}{mS/cm^2}\\
            $g_\text{syn}\equiv \ve$& \SIrange{0}{0.027}{mS/cm^2}\\
            $\alpha$ & 3\\
            $\beta$ & 2\\
            $\sigma_T$ & 0.8\\
            $V_\text{T}$& \SI{20}{mV}\\
            $I_\text{app}$& \SI{0.8}{\micro A/cm^2} (Figure \ref{fig:thal_examples1}) or \SI{0.6}{\micro A/cm^2} (Figure \ref{fig:thal_examples2})
        \end{tabular}
    \end{center}
\end{table}

The thalamic model is given by the equations,
\begin{align*}
    \frac{dV_i}{dt} &= -(I_\text{L}(V_i) + I_{\text{Na}}(V_i) + I_\text{K}(V_i) + I_\text{T}(V_i) +  \frac{g_\text{syn}}{N}\sum_{j=1}^N a_{ij} w_j(V_i-E_\text{syn})-I_\text{app})/C,\\
    \frac{dh_i}{dt} &= (h_\infty(V_i) - h_i)/\tau_h(V_i),\\
    \frac{dr_i}{dt} &= (r_\infty(V_i) - r_i)/\tau_r(V_i),\\
    \frac{dw_i}{dt} &= \alpha(1-w_i)/(1+\exp((V_i-V_\text{T})/\sigma_T)) - \beta w_i,
\end{align*}
where $i=1,\ldots,3$, and $a_{ii} = 0$ and $a_{ij}=1$ otherwise. Remaining equations are in Appendix \ref{a:thal} and all parameters are shown in Table \ref{tab:thal}. Given neuron $i$, the coupling term in the voltage variable $V_i$ is given by the average excitatory effect of the synaptic variables $w_j$ without self coupling. The synaptic conductance parameter $g_\text{syn}$ sets the overall coupling strength and is identical to the coupling strength parameter $\ve$.

\begin{figure}[ht!]
    \includegraphics[width=\textwidth]{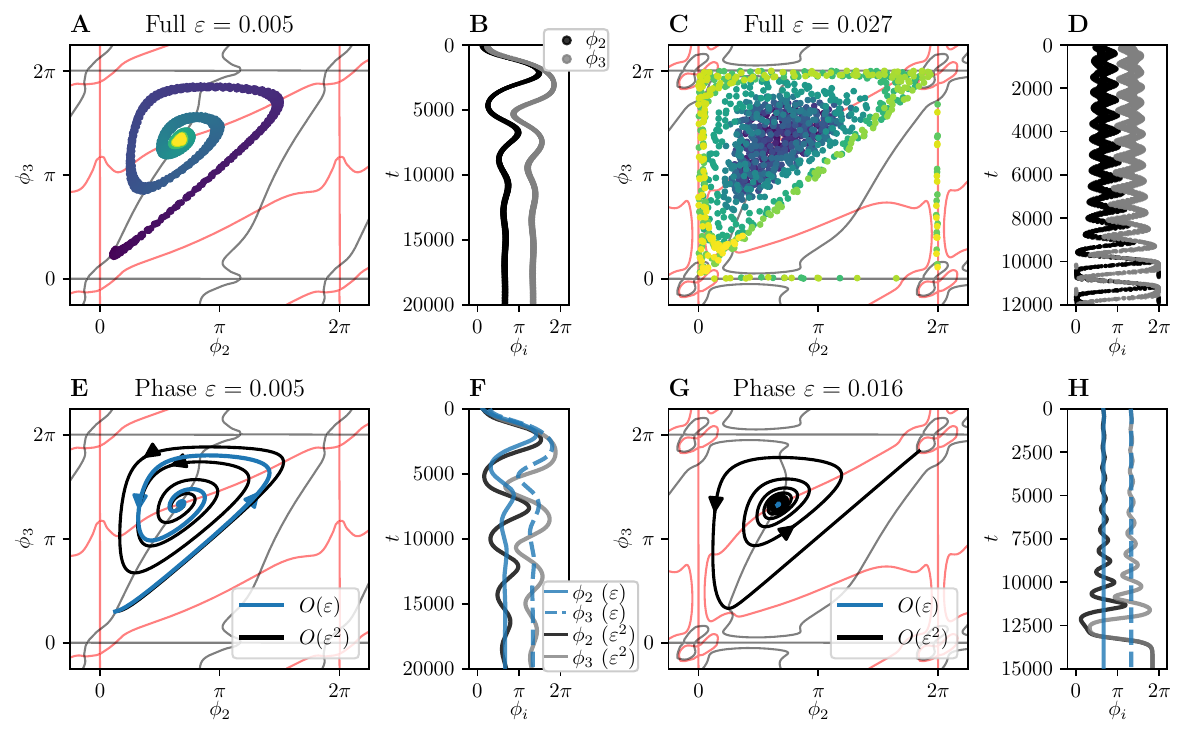}
    \caption{Comparison of the full (top row) and reduced (bottom row) thalamic models. All panels show the corresponding nullclines of the $O(\ve^2)$ reduced model. A: Phase difference estimate of the full model dynamics. Lighter shades indicate later times. E: The reduced models ($O(\ve)$ blue and $O(\ve^2)$ black) at the approximate corresponding coupling strength. Note that the $O(\ve)$ model remains at the splay state. Arrows indicate movement in forward time. B, F: Corresponding plots over time of the full model and reduced model, respectively. C, D, full and reduced model dynamics, respectively. D, H: corresponding plots over time of the full and reduced models, respectively. Parameters as in Table \ref{tab:thal} with $E_\text{syn}=\SI{0}{mV}$ and $I_\text{app}=\SI{0.8}{\micro A/cm^2}$. \ypb{We show every hundredth time point of the full model phase estimation}.}\label{fig:thal_examples1}
\end{figure}

We compare the reduced and full versions of the thalamic model in Figure \ref{fig:thal_examples1}, where the parameters are chosen as in Table \ref{tab:thal} with with $E_\text{syn}=\SI{0}{mV}$ and $I_\text{app}=\SI{0.8}{\micro A/cm^2}$. The top row shows phase estimates of the full model for $\ve=0.005$ (A) and $\ve=0.027$ (C), where lighter shades correspond to later times \yp{(see Appendix \ref{a:phase_estimate} for more details on the phase estimation of the full model)}. The bottom row shows the $O(\ve)$ (blue) and $O(\ve^2)$ (black) phase models exhibiting qualitatively similar phase dynamics at $\ve=0.005$ (E) and $\ve=0.016$ (G), respectively. Corresponding time traces are shown to the right of each portrait, e.g., panel B corresponds to A, and F corresponds to E.

\begin{figure}[ht!]
    \includegraphics[width=\textwidth]{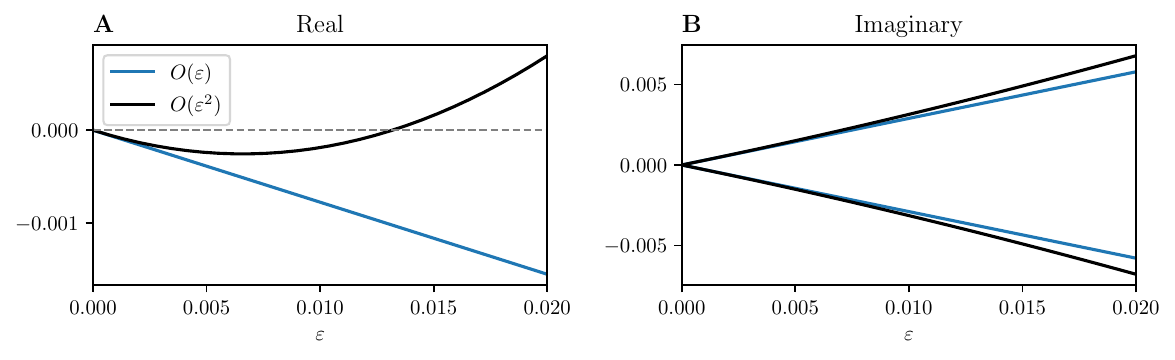}
    \caption{Real (A) and imaginary (B) parts of the eigenvalues of the Jacobian matrix evaluated at the splay state in the reduced thalamic models. Blue: $O(\ve)$, orange: $O(\ve^2)$, black: $O(\ve^3)$. \yp{The dashed line is provided for reference to highlight the loss of stability as $\ve$ increases in higher order accuracy models.} Parameters are identical to those used in Figure \ref{fig:thal_examples1}. }\label{fig:thal_eigs1}
\end{figure}

At $\ve=0.005$, the full model phase differences tend towards an asymptotically stable splay state when initialized near synchrony with phases $(\phi_2,\phi_3)=(0.4,1)$ (A,B, where $\phi_i \in [0,2\pi)$). With the same initial values, all $O(\ve)$ (blue) and $O(\ve^2)$ (black) phase models coincide with the full model at as expected (E,F).

For greater values of $\ve$, phase differences in the full model asymptotically tend towards a limit cycle oscillation (C,D) and the $O(\ve^2)$ reduced model tends towards synchrony. While the asymptotic dynamics differ, we nevertheless capture the loss of stability in the splay state (\yp{we checked the order $O(\ve^3)$ term, but found no improvement in the reduced solution}). Real and imaginary parts of the Jacobian evaluated at the splay state is shown in Figure \ref{fig:thal_examples1}.

\begin{figure}[ht!]
    \includegraphics[width=\textwidth]{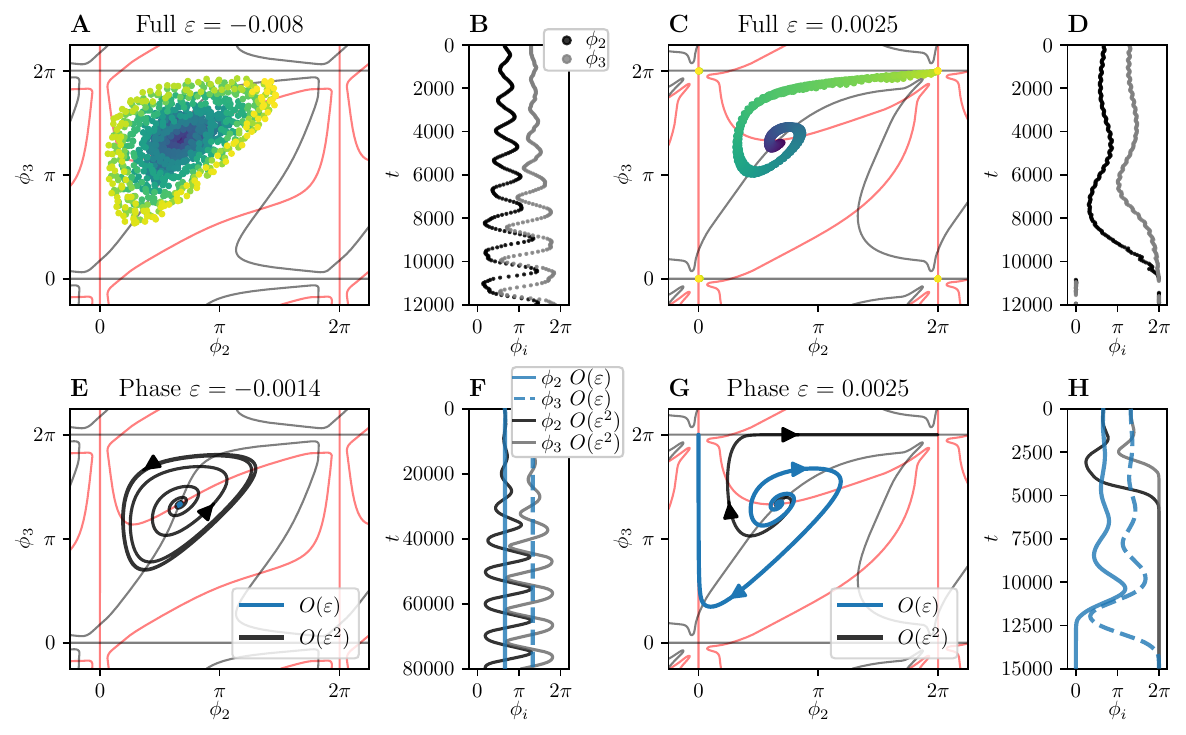}
    \caption{Comparison of the full (top row) and reduced (bottom row) thalamic models. All panels show the corresponding nullclines of the $O(\ve^2)$ reduced model. A: Phase difference estimate of the full model dynamics. Lighter shades indicate later times. E: The reduced models ($O(\ve)$ blue and $O(\ve^2)$ black) at the approximate corresponding coupling strength. Note that the $O(\ve)$ model remains at the splay state. Arrows indicate movement in forward time. B, F: Corresponding plots over time of the full model and reduced model, respectively. C, D, full and reduced model dynamics, respectively. D, H: corresponding plots over time of the full and reduced models, respectively. Parameters as in Table \ref{tab:thal} with $E_\text{syn}=\SI{-100}{mV}$ and $I_\text{app}=\SI{0.6}{\micro A/cm^2}$. \ypb{We show every hundredth time point of the full model phase estimation}.}\label{fig:thal_examples2}
\end{figure}

To further demonstrate the utility of our method, we show the phase reduction of the thalamic model for a different set of synaptic parameters: $E_\text{syn}=0$ and $\ve<0$. This choice is less biologically relevant because it corresponds to an excitatory chemical synapse with a negative conductance, but the goal of this example is to show that the reduced model can capture additional nonlinear dynamics in a model more complex than the CGL model.

\begin{figure}[ht!]
    \includegraphics[width=\textwidth]{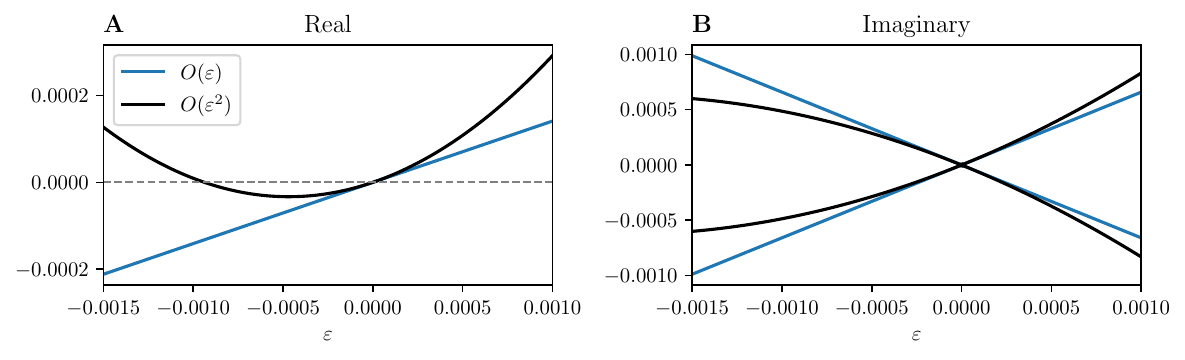}
    \caption{Real (A) and imaginary (B) parts of the eigenvalues of the Jacobian matrix evaluated at the splay state in the reduced thalamic models. Blue: $O(\ve)$, orange: $O(\ve^2)$, black: $O(\ve^3)$. \yp{The dashed line is provided for reference to highlight the change of stability as $\ve$ increases in higher order accuracy models.} Parameters are identical to those used in Figure \ref{fig:thal_examples2}.}\label{fig:thal_eigs2}
\end{figure}

Comparisons between the full and reduced versions of the thalamic model for this new parameter set are shown in Figure \ref{fig:thal_examples2}. The top row shows phase estimates of the full model for $\ve=-0.008$ (A) and $\ve=0.0025$ (C), where lighter shades correspond to later times \yp{(see Appendix \ref{a:phase_estimate} for more details on the phase estimation of the full model)}. The bottom row shows the $O(\ve)$ (blue) and $O(\ve^2)$ (black) phase models exhibiting qualitatively similar phase dynamics at $\ve=-0.0014$ (E) and $\ve=0.0025$ (G), respectively. Corresponding time traces are shown to the right of each portrait, e.g., panel B corresponds to A, and F corresponds to E.

The right column of Figure \ref{fig:thal_examples2} (panels C,D and G,H) serves as a sanity check, where $\ve>0$ puts us back in a biologically realistic regime. The full and reduced models all tend towards synchrony, however, the $O(\ve^2)$ model (black) captures the transient dynamics more accurately than the $O(\ve)$ model.

At $\ve=-0.008$, the full model phase differences exhibit a loss of stability in the splay state and the asymptotic dynamics tend towards a limit cycle (A,B). The $O(\ve^2)$ reduced model captures this behavior (E,F, black), whereas the $O(\ve)$ reduced model does not (E,F, blue).  Note the considerable nonlinearity in the phase difference dynamics as a function of $\ve$. Despite the small size of $\ve$, the $O(\ve)$ and $O(\ve^2)$ dynamics differ substantially.

This example is affected by a nearby saddle-node on an invariant cycle (SNIC) bifurcation, which occurs just below the applied current value of $I_\text{app}=\SI{0.6}{\micro A/cm^2}$, and reminds us that ``small'' $\ve$ is relative. The first example of the thalamic model (Figure \ref{fig:thal_examples1}) uses much greater values of $\ve=0.005$ and $\ve=0.027$, the latter being an order of magnitude greater than in the current example (Figure \ref{fig:thal_examples2}). {The proximity to a SNIC also highlights the shortcomings with our second assumption, where we use first order averaging. The reciprocal of our period is $1/T=1/\SI{44}{ms}\approx 0.023$, which places an approximate upper bound on the coupling strength $\ve$, which must be much smaller than $1/T$ \cite{maggia2020higher}.}

Nevertheless, we can compute changes in the stability of the splay state as in the previous examples. The real and imaginary parts of the eigenvalues of the Jacobian matrix evaluated at the splay state is shown in Figure \ref{fig:thal_eigs2}. The $O(\ve^2)$ model captures the loss in stability while the $O(\ve)$ model does not.


\section{Discussion}\label{sec:discussion}

In summary, we derived coupling functions that capture higher-order $N$-body interactions \yp{while allowing for nontrivial effects from slowly decaying Floquet modes}. While we did not consider heterogeneity, the formulation allows for the vector fields to be entirely different, so long as \yp{there exists a parameter in each system that can bring the oscillators to the same period, or alternatively, so long as} the oscillator periods are similar in the absence of coupling. We applied our method to two systems, the CGL model and a thalamic neuron model. We found that higher-order terms were necessary to reproduce the dynamics of the original system. In the CGL model, even though the $O(\ve^2)$ reduced model contained 3-body interactions, it was the $O(\ve^3)$ reduced model that captured additional nonlinearities, suggesting that in general, $N$-body interactions alone are not always sufficient to reproduce full model dynamics. In the thalamic model, we considered two examples, the first at $I_\text{app}=\SI{0.8}{\micro A/cm^2}$ $E_\text{syn}=-100$ and the second at $I_\text{app}=\SI{0.6}{\micro A/cm^2}$ $E_\text{syn}=0$. In the first example, we captured the loss in stability of the splay state using the $O(\ve^2)$ model. In the second example, we exploring beyond a biologically relevant parameter regime and observed limit cycle behavior in the phase difference of the full model for $g_\text{syn}\equiv \ve <0$, which the $O(\ve^2)$ reduced model captured. We noted that the $\ve$ values between the full and reduced models differed more than previous examples to capture similar behaviors. This difference is likely due to the existence of a SNIC bifurcation just below the parameter value $I_\text{app}=\SI{0.6}{\micro A/cm^2}$.

Our method is both a generalization of existing methods that consider higher-order phase-isostable interactions and a general framework from which to study higher-order effects. For example, a higher-order reduced model is derived using the Haken-Kelso-Bunz (HKB) equation in \cite{mckinley2021third}. The higher-order terms are the lowest-order Fourier terms of our $\mathcal{H}$ functions, thus the same questions of existence can be answered with our method and further explored with additional Fourier terms and multi-body interactions. Larger networks of the HKB equation that consider interactions well beyond dyadic \cite{zhang2019connecting} fit comfortably within the limitations of our method (see Section \ref{sec:limits} below for details). Similarly, there is no restriction to applying our method to questions of coordinated movement, e.g., \cite{kelso2021unifying}, or studies of coupled population dynamics \cite{milne2021coupled}.

Our method may aid in addressing questions of synchrony and phase-locking in general  finite populations of coupled oscillators with heterogeneity where order parameters are typically used. For example, the heterogeneous systems and coupling functions considered in \cite{alderisio2016entrainment} can not exhibit synchrony and a ``bounded synchronization'' measurement \cite{hill2008global} is necessary. Our method could provide a far more detailed understanding of the bounded synchronization state alongside other possible phase-locked states. Moreover, similar questions could be asked in much more realistic and complex neurobiological models.

Two recent results are most relevant to our work. The first is \cite{mau2023high}, where a phase-isostable description of coupled oscillators is derived for heterogeneous, planar oscillators. Their assumptions and handling of the isostable coordinate(s) are similar to ours (including earlier work \cite{park2021high}). In particular, their heterogeneous assumption is equivalent to ours in the following sense. If we let $\mu=1,\ldots,M$ be the oscillator index for $M$ oscillators, then we can directly transform our framework into theirs by taking $\delta_\mu \mathcal{Z}_\mu(\theta_\mu,\psi_\mu)\cdot Q_i(\theta_\mu,\psi_\mu)$ to lowest order in $\ve$, which is some constant $\delta_\mu \hat{d}$, say, and setting $1+\delta_\mu \hat{d} = \omega_\mu$. Then our framework can be used to explore phase-locking and phase drift as a function of coupling strength and forcing frequency in the same manner. A key difference is that our framework allows oscillators of arbitrarily high dimension (not to mention that we publicly share our numerical methods under an open-source license \cite{parkgithub}).

Finally, the remarkable work by Nicks et al \cite{nicks2024insights} warrants special attention. They use both phase and isostable coordinates to derive conditions for the existence and stability of phase-locked states in networks of coupled oscillators. The isostable dynamics are kept explicit, which confers some advantages that our method may not. Indeed, the choice of ``eliminating'' the isostable dynamics in our work inevitably reduces the accuracy of our approximation, because we ignore potentially important transient effects. \ypb{Moreover, analytical calculations become rapidly cumbersome for lower-order terms for even very simple models such that numerics become necessary beyond order 3 or 4.} Their work highlights the importance of choosing methods with compatible assumptions when applying phase-isostable methods to ones own models. It may very well be worth increasing the accuracy of a reduced system in exchange for an increase in dimension. On the other hand, our method is useful if only pure phases are desired, while including at least some of the corrective effects of isostable dynamics.

\subsection{Limitations}\label{sec:limits}

We begin with limits related to our implementation. If the $\mathcal{H}$ functions have \yp{sparse Fourier modes}, then a Fourier truncation can be used to greatly reduce the time complexity and memory requirements of our method (see Appendix \ref{a:fourier}). In particular, knowledge of the exact types of functions that appear for each order is a significant part of this computational efficiency. However, the current implementation has only up to order $O(\ve^3)$ implemented for the Fourier method. While it is clear that there is a pattern in the types of separable and non-separable functions that appear in the Fourier terms as a function of higher orders, we have not precisely determined a formula for this pattern at the time of this writing. Once the pattern is fully understood, it will be possible to determine significantly higher-order interaction terms using the Fourier truncation.

Limitations related to the reduction method outside of our implementation center around the two key assumptions that make the derivation of this method possible. First, we assume that the timescale of phase $\theta_i-t$ (i.e., the phase equation after subtracting the moving frame) differs sufficiently from the timescale of the functions $p_i^{(k)}$ (the expansion terms of the isostable coordinate $\psi_i$), such that $\theta_i$ terms can be taken out of a time integral. This assumption is reasonable for at least moderate values of $\ve$ up to $\ve=0.06$ in our examples, where the phase difference variables $\phi_i$ tend to vary relatively slowly. However, additional work must be done to more carefully examine this assumption for use in greater stronger coupling strenghts $\ve$.

Second, we use \textit{first order} averaging, which is technically valid for small $\ve$ comparable to those used in weak coupling theory. This limitation is especially apparent in the last example, where the thalamic model is near a SNIC bifurcation and the reciprocal of the period {($1/\SI{44}{ms}\approx 0.023$) places an approximate upper bound on the coupling strength $\ve$, as $\ve$ must be much smaller than $1/T$ \cite{maggia2020higher}}. This example may benefit from higher-order averaging methods \cite{llibre2014higher,maggia2020higher} could be used. In addition, we have observed phase drift in the full model (data not shown) in a manner that may not be possible to capture in the current formulation. For example, with $N=3$ homogeneous oscillators and some values of $\ve$, two oscillators synchronize and the third exhibits a phase drift, effectively resulting in a 2 oscillator system with a drift in the remaining phase difference. In our formulation, a single phase difference equation can not exhibit drift without heterogeneity. \yp{This discrepancy may be due to ignoring transients in the isostable coordinates -- if we were to include explicit isostable dynamics such as in \cite{nicks2024insights}, this behavior may be captured.}

\section{Acknowledgments}

The authors acknowledge support under the National Science Foundation Grant No. NSF-2140527 (DW). \ypb{The authors acknowledge Reviewer \#2's diligence toward checking many important details in this text -- their questions and suggestions significantly improved its content and clarity}

\appendix

\section{Numerical Details for Computing Higher-Order Interaction Functions}\label{a:numerics}

We briefly discuss three numerical methods to calculate $\mathcal{H}$ functions.

\subsection{Option 1: Averaging in Terms of Fourier Coefficients}\label{a:fourier}
If the expansion terms $g_i^{(k)}$, $\mathcal{Z}_i^{(k)}$ and $\mathcal{I}_i^{(k)}$ have sparse Fourier modes, it may be advantageous to calculate the $\mathcal{H}$ functions in terms of Fourier coefficients. Given order $k$, define $\bm{\Theta}=(\theta_1,\ldots,\theta_N)$ and $\bm{1}=(1,\ldots,1)$. \yp{Suppose that we wish to obtain a $\mathcal{H}_i^{(k)}$ function, so we apply first-order averaging to some order $k$ function $f^{(k)}$ for a given system of $N$ oscillators}:
\begin{equation*}
    \int_0^T \bar f^{(k)}(\theta_1+t,\ldots,\theta_N+t)\, \mathrm{d}t.
\end{equation*}
\yp{The integrand $f^{(k)}$ is simply a placeholder for any of the terms that appear in the integrand in the calculation of $\mathcal{H}$ functions, e.g., for $k=1$, $f^{(1)}$ could be any integrand term here:
\begin{align*}
    \sum_{j_1,j_2} \mathcal{H}_{ij_1j_2}(\xi_1,\xi_2) =&\sum_{j_1,j_2}  \frac{a_{ij_2}}{T}\int_0^T Z_i^{(0)}(s)\cdot v_{ij_1}^{(1,0)}(s,\xi+s)  g_{ij_2}(s,\xi+s)\dd s\\
    &+\sum_{j_1,j_2}  \frac{a_{j_1j_2}}{T}\int_0^T Z_i^{(0)}(s)\cdot v_{ij_1}^{(1,0)}(s,\xi_1+s)  g_{j_1j_2}(\xi_1+s,\xi_2+s)\dd s\\
    &+\sum_{j_1,j_2}  \frac{a_{ij_2}}{T}\int_0^T Z_i^{(0)}(s)\cdot K_{ij_1}^{(0)}(s,\xi_1+s)  g_{ij_2}(s,\xi_2+s)\dd s.
\end{align*}}
Consider the $N$-dimensional Fourier expansion of the integrand:
\begin{equation*}
     \int_0^T \sum_{\bm{m}\in\mathbb{Z}^N} c_{j,k,\bm{m}} e^{i \bm{m}\cdot(\bm{\Theta}+\bm{1}t)}\, \mathrm{d}t,
\end{equation*}
where $c_{j,k,\bm{m}}$ are the Fourier coefficients given oscillator $j$ and order $k$, and $\bm{m} = (m_1,m_2,\ldots,m_N)$.
We simplify this integral using straightforward integral properties and orthogonality of the Fourier basis:
\begin{equation}\label{eq:fourier_avg}
    \begin{split}
        \int_0^T \sum_{\bm{m}\in\mathbb{Z}^N} c_{i,k,\bm{m}} e^{i \bm{m}\cdot(\bm{\Theta}+\bm{1}t)}\, \mathrm{d}t &=  \sum_{\bm{m}\in\mathbb{Z}^N}  \int_0^T c_{i,k,\bm{m}} e^{i \bm{m}\cdot(\bm{\Theta}+\bm{1}t)}\, \mathrm{d}t\\
        &= \sum_{\bm{m}\in\mathbb{Z}^N} c_{i,k,\bm{m}} e^{i \bm{m}\cdot\bm{\Theta}}\int_0^T  e^{i\bm{m}\cdot\bm{1}t}\, \mathrm{d}t\\
        &= \sum_{\bm{m}\cdot\bm{1}=0} c_{i,k,\bm{m}} e^{i \bm{m}\cdot\bm{\Theta}}.
    \end{split}
\end{equation}
That is, the only integral terms that remain are those such that the index vector $\bm{m}$ is orthogonal to the vector $\bm{1}$, and the integral of those surviving terms trivially evaluate to the scalar 1 \yp{(this calculation is a special case of \cite{ermentrout1981})}. Thus, the question of computing the average in time is simply a matter of computing the set of Fourier coefficients and extracting the relevant subset. If the number of oscillators $N$ and mesh size $K$ are large, then the desired averaged function in its final form can be taken as above, otherwise the desired functions can be found by the inverse Fourier transform.

\paragraph{Claim: The right-hand side of \eqref{eq:fourier_avg} depends on $N-1$ variables}

Without loss of generality, suppose $j=1$. Now consider the change of variables $s=\theta_1+t$ and recall that $\phi_i = \theta_i-\theta_1$. Then rewriting \eqref{eq:fourier_avg} yields
\begin{align*}
    \sum_{\bm{m}\in\mathbb{Z}^N} c_{1,k,\bm{m}} \int_0^T e^{i \bm{m}\cdot(s,\phi_2+s,\ldots,\phi_N+s)}\, \mathrm{d}s &= \sum_{\bm{m}\in\mathbb{Z}^N} c_{1,k,\bm{m}} \int_0^T e^{i\left[ m_1 s+m_2(\phi_2+s)+\cdots+m_N(\phi_N+s)\right]}\, \mathrm{d}s\\
    &= \sum_{\bm{m}\in\mathbb{Z}^N} c_{1,k,\bm{m}} e^{i(m_2 \phi_2+\cdots+m_N \phi_N)}\int_0^T e^{i s \bm{m}\cdot \bm{1}}\, \mathrm{d}s\\
    &= \sum_{\bm{m}\cdot \bm{1} = 0} c_{1,k,\bm{m}} e^{i(m_2 \phi_2+\cdots+m_N \phi_N)},
\end{align*}
where the last lines uses the orthogonality of the Fourier basis to evaluate the integral. Thus, we may easily evaluate \eqref{eq:fourier_avg} for $N-1$ phase differences by evaluating any one coordinate at zero.

While \eqref{eq:fourier_avg} is computationally efficient, we once again find that evaluating the underlying $N$-dimensional function to compute the Fourier coefficients of \eqref{eq:fourier_avg} requires significant amounts of memory. For example, evaluating an $N$ dimensional function on a mesh size of $K$ again requires $N^K$ points prior to taking the Fourier coefficient. So we seek an additional reduction in memory.

\subsection{Option 2: Using an ODE Solver as an Adaptive Mesh}\label{a:ode_mesh}
In cases where the response functions ($\mathcal{Z}$ and $\mathcal{I}$) and Floquet eigenfunctions have derivatives that greatly exceed their magnitudes, it is not possible to use only a small number of Fourier coefficients. Indeed, the thalamic model has a relatively large number of nontrivial Fourier modes for each order. For a uniform mesh, the number of points in the time-average integral grid $K$ exceeds $\num{1e7}$, which is difficult to compute efficiently. Parallelizing this problem is further hindered by memory constraints.

Because the sharp peaks in the response functions and eigenfunctions tend to be in small regions of phase space, an adaptive mesh greatly reduces the number of integration points $K$. The most straightforward method is to use an ODE solver by rephrasing the time-average integration as an initial value problem. Given $k$, and $\theta_1,\ldots,\theta_N$, rewrite
\begin{equation*}
    \begin{split}
        \mathcal{H} &= \frac{1}{T}\int_0^T \bar h_j^{(k)}(\theta_1+t,\ldots,\theta_N+t)\, \mathrm{d}t
    \end{split}
\end{equation*}
as
\begin{equation*}
    \begin{split}
        T \frac{d}{dt}\mathcal{H}(t)&= \bar h_j^{(k)}(\theta_1+t,\ldots,\theta_N+t), \quad \mathcal{H}(0) = 0.
    \end{split}
\end{equation*}
Then the desired time-average is given by $\mathcal{H}(T)$. If needed, the mesh is provided by the ODE solver. We use the Python \cite{10.5555/1593511} implementation of LSODA.

\subsection{Option 3: Brute Force and Parallelization}

If all else fails, it is possible to brute force calculations of the $\mathcal{H}$ functions, especially if the network has $N\leq 3$ oscillators with only a few lower-order terms and a coarse mesh. Our Python implementation will attempt to use CPU parallelization, and if available, CUDA parallelization. Because the memory requirements grow exponentially as a function of mesh size, mesh sizes are typically restricted to 500 points for roughly 16GB of RAM. Running the code on a cluster with more CPUs is recommended (instructions are included in the repository with sample scripts).

\section{Thalamic Model}\label{a:thal}

The remaining equations for the Thalamic model are
\begin{align*}
    I_\text{L}(V) = g_\text{L} (V-E_L), &\quad I_{\text{Na}} = g_\text{Na} h m_\infty^3(V)(V-E_\text{Na}),\\
    I_\text{K} = 0.75 g_\text{K}(1-h)^4(V-E_\text{K}), &\quad I_\text{T} = g_\text{T} r p_\infty^2(V)  (V-E_\text{T}),
\end{align*}
and
\begin{align*}
    a_h(V) = 0.128 \exp(-(V+46)/18),&\quad b_h(V) = 4/(1+\exp(-(V+23)/5)),\\
    m_\infty(V) = 1/(1+\exp(-(V+37)/7)),&\quad h_\infty(V) = 1/(1+\exp((V+41)/4)),\\
    r_\infty(V) = 1/(1+\exp((V+84)/4)),&\quad p_\infty(V) = 1/(1+\exp(-(V+60)/6.2)),\\
    \tau_h(V) = 1/(a_h(V)+b_h(V)),&\quad \tau_r(V) = 28 +\exp(-(V+25)/10.5).
\end{align*}

\section{Coupling Function Expansions}\label{a:expansion}
Recall our original coupled system \eqref{eq:odes},
\begin{equation*}
    \dot X_i = F_i(X_i) + \delta_i \yp{Q}_i(X_i) + \ve \left[ \sum_{j=1}^N a_{ij} G_{ij}(X_i,X_j) \right], \quad i=1,2,\ldots,N,\\
\end{equation*}
Here we provide a high-level description of the $\ve$-expansion of the coupling functions $G_{ij}$. First, fix $i$ and $j$ and consider an arbitrary $m$th coordinate $G_m$ of $G_{ij}$. Recall that we expand this function in the following manner,
\begin{equation}\label{eq:g_exp_a}
    G_{m}(\Lambda + \Delta \Xi)= G_m^{(0)}(\Lambda)+ G_m^{(1)}(\Lambda)\Delta \Xi+ \sum_{k=2}^\infty \frac{1}{k!}\left[ \stackrel{k}{\otimes} \Delta \Xi^\tr\right] \vc\left(G_{m}^{(k)}(\Lambda)\right),
\end{equation}
where $\Delta \Xi = [\Delta x^\tr,\Delta y^\tr]^\tr\in \mathbb{R}^{n_i+n_j}$, $\Lambda$ is purely a function of $\theta_i$ and $\theta_j$, and
$\Delta x \approx \psi_i g_i^{(1)}(\theta_i) + \psi_i^2 g_i^{(2)}(\theta_i) + \cdots$, with $\Delta y$ defined similarly. Writing $\Delta x = [\Delta x_1,\ldots,\Delta x_{n_i}]^T$, each $\Delta x_\ell$ is a polynomial in $\psi_i$ by definition (likewise for $\Delta y_\ell$). It follows that $\stackrel{1}{\otimes} \Delta \Xi^\tr\equiv  \Delta \Xi$ only contains polynomials in $\psi_{i}$ in the first $n_i$ elements and polynomials in $\psi_j$ in the last $n_j$ elements. Thus, the term $G_m^{(1)}(\Lambda)\Delta \Xi$, which is equivalent to a dot product, is a scalar that consists of the sum of a polynomial in $\psi_i$ with a sum of a polynomial in $\psi_j$. The next term in the expansion includes the cross-multiplication of the two polynomials:
\begin{align*}
    \stackrel{2}{\otimes} \Delta \Xi^\tr &\equiv [\Delta x_1 \Delta \Xi^\tr, \Delta x_2 \Delta \Xi^\tr, \cdots \Delta x_{n_i} \Delta \Xi^\tr, \Delta y_1 \Delta \Xi^\tr, \cdots, \Delta y_{n_j} \Delta \Xi^\tr]\\
    &\equiv [[\Delta x_1^2, \Delta x_1 \Delta x_2, \ldots, \Delta x_1 \Delta x_{n_i}, \Delta x_1 \Delta y_1,\ldots \Delta x_1 \Delta y_{n_j}],\ldots,\\
    &\quad\quad[\Delta x_{n_i}\Delta x_1, \Delta x_{n_i}\Delta x_2, \ldots, \Delta x_{n_i}^2, \Delta x_{n_i}\Delta y_1,\ldots \Delta x_{n_i} \Delta y_{n_j}],\ldots,\\
    &\quad\quad[\Delta y_1\Delta x_1, \ldots, \Delta y_1 \Delta x_{n_i}, \Delta y_{1}^2,\ldots,\Delta y_1 \Delta y_{n_j}],\ldots,\\
    &\quad\quad[\Delta y_{n_j}\Delta x_1, \ldots, \Delta y_{n_j} \Delta x_{n_i}, \Delta y_{n_j}\Delta y_1,\ldots,\Delta y_{n_j}^2]],
\end{align*}
where the square brackets only help organize the terms. This vector has the shape $1\times (n_i + n_j)^2$ and contains polynomials of the form
\begin{equation*}
    (\psi_i c_1 + \psi_i^2 c_2 +\cdots)^\ell (\psi_j d_1 + \psi_j^2 d_2 +\cdots)^{(2-\ell)},
\end{equation*}
for some arbitrary scalars $c_k$, $d_k$, $k\in\mathbb{N}^+$, and for $0\leq \ell \leq 2$. Thus, the scalar term
\begin{equation*}
    \left[\stackrel{2}{\otimes} \Delta \Xi^\tr\right]\vc\left(G_m^{(2)}(\Lambda)\right)
\end{equation*}
contains a sum of polynomials in $\psi_i$ and $\psi_j$ with minimum order 2. Indeed, continuing this argument for arbitrary $k$ yields polynomials in $\psi_i$ and $\psi_j$ of minimum degree $k$ in the scalar term
\begin{equation*}
    \left[\stackrel{k}{\otimes} \Delta \Xi^\tr\right]\vc\left(G_m^{(k)}(\Lambda)\right).
\end{equation*}
Since $m$ was chosen arbitrarily, it follows that each coordinate of $G_{ij}$ contains the same types of polynomials, and we can express each term in \eqref{eq:g_exp_a} as polynomials with vector coefficients. We assume that the polynomial terms have been collected and let $v_{ij}^{(\ell_1,\ell_2)}(\theta_i,\theta_j)$ be the vector coefficient of the term $\psi_i^{\ell_1}\psi_j^{\ell_2}$. Then  we may express the expansion of $G_{ij}$ in terms of $\psi_i$ and $\psi_j$:
\begin{align*}
    G_{ij}(\Lambda(\theta_i,\theta_j) + \Delta \Xi(\theta_i,\theta_j,\psi_i,\psi_j)) =  \sum_{k=0}^\infty \sum_{\ell=0}^k \psi_i^\ell \psi_j^{(k-\ell)} v_{i,j}^{(\ell,k-\ell)} (\theta_i,\theta_j).
\end{align*}
Finally, we make the substitution $\psi_i = \ve p_i^{(1)} + \ve^2 p_i^{(2)} + \cdots$, yielding the following terms ordered in powers of $\ve$:
\begin{equation*}
    \begin{split}
        O(\ve^0)&:v_{ij}^{(0,0)}\\
        O(\ve^1)&:p_i^{(1)}v_{ij}^{(1,0)} + p_j^{(1)} v_{ij}^{(0,1)}\\
        O(\ve^2)&:p_i^{(2)}v_{ij}^{(2,0)} + p_j^{(2)}v_{ij}^{(0,2)} + (p_i^{(1)})^2v_{ij}^{(2,0)} + (p_j^{(1)})^2 v_{ij}^{(0,2)} + p_i^{(1)}p_j^{(1)} v_{ij}^{(1,1)}\\
        O(\ve^3)&:p_i^{(3)} v_{ij}^{(3,0)} + p_j^{(3)}v_{ij}^{(0,3)} + [p_i^{(1)}p_j^{(2)} + p_i^{(2)}p_j^{(1)}]v_{ij}^{(1,1)} + (p_i^{(1)})^3 v_{ij}^{(3,0)} + (p_j^{(1)})^3 v_{ij}^{(0,3)}\\
        &\vdots.
    \end{split}
\end{equation*}
Thus, we have a more explicit form for the $K_{ij}^{(\ell)}$ $\ve$-expansion terms of $G_{ij}$, assuming that $v_{ij}^{(\ell_1,\ell_2)}$ is known. The calculation of $v_{ij}^{(\ell_1,\ell_2)}$ is handled automatically by a symbolic package.

\subsection{Thalamic Model}

\begin{align*}
    K_{ij}^{(0)} &= \frac{1}{C}[a_{ij}E_{\text{syn}}w_j - a_{ij}v_iw_j,0,0,0]^\tr,\\
    K_{ij}^{(1)} &= \frac{1}{C}[a_{ij}E_{\text{syn}}g_{w_j}^{(1)}p_j^{(1)} - a_{ij}v_ig_{w_j}^{(1)}p_j^{(1)} - a_{ij}w_jg_{v_i}^{(1)}p_i^{(1)},0,0,0]^\tr,\\
    K_{ij}^{(2)} &= \frac{1}{C}
    \begin{bmatrix}
    \begin{array}{l}a_{ij}E_{\text{syn}}g_{w_j}^{(1)}p_j^{(2)} + a_{ij}E_{\text{syn}}g_{w_j}^{(2)}(p_j^{(1)})^2 - a_{ij}v_ig_{w_j}^{(1)}p_j^{(2)} - a_{ij}v_ig_{w_j}^{(2)}(p_j^{(1)})^2\\
         - a_{ij}w_jg_{v_i}^{(1)}p_i^{(2)} - a_{ij}w_jg_{v_i}^{(2)}(p_i^{(1)})^2 - a_{ij}g_{v_i}^{(1)}g_{w_j}^{(1)}p_i^{(1)}p_j^{(1)} \end{array},0,0,0
    \end{bmatrix}^\tr.
\end{align*}

\section{Phase Estimation}\label{a:phase_estimate}

We briefly describe the phase estimation method used in the paper (which is very similar to the estimation done in \cite{park2016weakly}). Consider a model with state variables $x_1,\ldots,x_n$, and suppose that we have saved a $T$-periodic limit cycle trajectory (at $\ve=0$) to some array $[y_1,\ldots,y_n]$. Then for a given simulation, we can define the phase to be a point $\theta\in[0,T)$ that minimizes
\begin{equation*}
    \text{dist}(x_1(t)-y_1(\theta),\ldots, x_n(t)-y_n(\theta)),
\end{equation*}
where
\begin{equation*}
    \text{dist}(\Delta x_1,\ldots, \Delta x_n) := \sqrt{(\Delta x_1)^2 + \cdots + (\Delta x_n)^2}.
\end{equation*}
By simulating nondimensionalized versions of the equations, we need not normalize by the variance as in \cite{park2016weakly}. This brute-force method is remarkably fast with appropriate vectorization.

\end{document}